%% file: main_arxiv.tex
\documentclass[a4paper]{article}
\usepackage{Odyssey2020}
\usepackage{epsfig,amssymb,amsmath,rotating}
\ninept
\usepackage[justification=centering]{caption}
\input{preambles}
\usepackage[labelsep=colon]{caption}
\usepackage{dblfloatfix}    
\usepackage[latin1]{inputenc}
\usepackage{amsmath}
\usepackage{amsfonts}
\usepackage{amssymb}
\usepackage{epstopdf}
\usepackage{mwe}
\usepackage{multirow}
\usepackage[export]{adjustbox}
\usepackage{algorithm}
\usepackage{kantlipsum}
\usepackage{xpatch,xcolor}
\usepackage{graphicx}
\usepackage{graphics}
\usepackage{afterpage}
\usepackage{balance}
\usepackage{xcolor}
\usepackage{amstext}
\usepackage{xspace}
\makeatletter
\def\BState{\State\hskip-\ALG@thistlm}
\makeatother
\title{An explainability study of the constant Q cepstral coefficient \\ spoofing countermeasure for automatic speaker verification}

\name{Hemlata Tak, Jose Patino, Andreas Nautsch, Nicholas Evans and Massimiliano Todisco}

\address{EURECOM, Sophia Antipolis, France\\
{\small \tt lastname@eurecom.fr} }
\begin{document}
\maketitle

\begin{abstract}
Anti-spoofing for automatic speaker verification is now a 
well established area of research, with three competitive challenges having been held in the last 6 years.
A great deal of research effort over this time has been invested into the development of front-end representations tailored to the spoofing detection task.
One such approach known as constant Q cepstral coefficients (CQCCs) have been shown to be especially effective in detecting attacks implemented with a unit selection based speech synthesis algorithm.
Despite their success, they largely fail in detecting other forms of spoofing attack where more traditional front-end representations give substantially better results.
Similar differences were also observed in the most recent, 2019 edition of the ASVspoof challenge series.
This paper reports our attempts to help explain these observations. 
The explanation is shown to lie in the level of attention paid by each front-end to different sub-band components of the spectrum.
Thus far, surprisingly little has been learned about what artefacts are being detected by spoofing countermeasures.
Our work hence aims to shed light upon signal or spectrum level artefacts that serve to distinguish different forms of spoofing attack from genuine, bone fide speech. 
With a better understanding of these artefacts we will be better positioned to design more reliable countermeasures.

\end{abstract}

\section{Introduction}
Automatic speaker verification (ASV) technology is ubiquitous nowadays, being applied to user authentication in an increasingly diverse array of applications from smart-home technology to telephone banking to health-sector applications~\cite{hansen2015speaker,kinnunen2010overview}.
Despite its success, there are justified concerns surrounding the vulnerabilities of ASV to spoofing, namely manipulation from inputs specially crafted to deceive the system and provoke false acceptances.
The community has responded to this threat with spoofing countermeasures that aim to detect and deflect such attacks.
The effort has been spearheaded through the ASVspoof initiative which emerged from the first special event on the topic held at INTERSPEECH in 2013~\cite{EURECOM+4018,evans2013spoofing}.

The first edition of ASVspoof, held in 2015~\cite{wu2015asvspoof}, showed that some voice conversion and speech synthesis spoofing attacks can be detected with ease.  
The infamous S10 unit selection speech synthesis algorithm,\footnote{\url{http://mary.dfki.de/}} however, was shown to present substantial difficulties.  
The constant Q cepstral coefficient (CQCC) front-end~\cite{todisco2016new,todisco2017constant}, introduced subsequently in 2016, was the first solution capable of detecting the S10 attack, reducing equal error rates (EERs) by some $72\%$ relative to the state of the art at the time. 
While better performing solutions have since emerged, the CQCC front-end remains popular.
Over half of ASVspoof 2019 participants (26 out of 48) used CQCC representations as a component of their submission.
This includes a number of top-performing systems, e.g.~\cite{chettri2019ensemble, alluri2019iiit,yang2019sjtu,lavrentyeva2019stc}.

CQCC features are, however, not a \emph{silver bullet}.
Even if near-to-zero EERs can be obtained for some attacks, for others results are poor. 
Among those in the ASVspoof 2015 database, for the S8 tensor-based approach to voice conversion~\cite{saito2011one}, EERs obtained using CQCCs are substantially higher than those obtained with more conventional linear frequency cepstral coefficient (LFCC) features.
These observations have led us to ask ourselves \emph{why}?  
What could account for these performance differences and what can be learned from the explanation?

The answer lies in the artefacts that characterise spoofing attacks for which each front-end excels. 
Artefacts are signatures or tell tale signs in spoofed speech that are left behind by voice conversion and speech synthesis algorithms.
They hence provide the means to distinguish between bona fide and spoofed speech.
While end-to-end deep learning architectures, which operate directly upon the signal and which have the capacity to identify such artefacts automatically, may diminish the need to understand what makes one representation outperform another, explainability is \emph{always} an issue of scientific interest.

While we have learned a great deal about the vulnerabilities of ASV to spoofing and the pattern recognition architectures that function well in detecting spoofing attacks, we have probably learned surprising little about what characterises spoofing attacks at the signal, spectrum or feature level. 
An understanding at this level would surely place us in a better position to design better performing countermeasures capable of detecting a broader range of attacks with greater reliability.
The same understanding might also provide some assurance that the artefacts being detected are in fact indicative of spoofing attacks rather than database collection and pre-processing procedures.

The work presented in this paper hence revisits CQCC features and attempts to show why they are effective in detecting some attacks but less effective in detecting others. 
In particular, we are interested in shedding light upon what is being detected in the signal, i.e.\ the artefacts that characterise a spoofing attack. The paper is organised as follows.  Section~2 presents a brief review of the CQCC front-end.  Section~3 describes a sub-band approach to examine where information relevant to spoofing detection is located in the spectrum.  Section~4 presents countermeasure performance and sub-band analysis results for the ASVspoof 2019 database.  Section~5 describes the impact of variable spectro-temporal resolution whereas Section 6 presents additional experiments designed to validate our findings.  A discussion of these is presented in Section~7, together with some directions for future work.

\section{Constant Q cepstral coefficients}
This section describes the motivation behind the development of constant Q cepstral coefficients (CQCCs).  It shows how CQCC features are derived starting with the constant Q transform (CQT), describes differences to spectro-temporal decomposition with the discrete Fourier transform (DFT) and presents comparative results for the ASVspoof 2015 database.

\subsection{Motivation}
\begin{table*}[!t]
	\centering
	\caption{EERs (\%) for the ASVspoof 2015 database, evaluation partition. Results for GMM-CQCC and GMM-LFCC baseline systems, reproduced from~\cite{todisco2016new}.  Results for two attacks (S8 and S10) show stark differences in performance for each system.}
	\setlength\tabcolsep{6pt}
	\begin{tabular}{ *{11}{c}}
		\hline
		System	&S1 &S2 & S3 & S4 & S5  &S6 &S7 &\textbf{S8} &S9 &\textbf{S10}\\ 
		\hline\hline
		GMM-CQCC& 0.005& 0.106& 0.000& 0.000& 0.130  &0.098& 0.064 &\color{purple}\textbf{1.033}& 0.053 &\color{blue}\textbf{1.065}\\		
		\hline 
		GMM-LFCC&0.027& 0.408& 0.000& 0.000& 0.114& 0.149& 0.011& \color{blue}\textbf{0.074}& 0.027& \color{purple}\textbf{8.185}\\
		\hline
	\end{tabular}
	\label{Tab:2015 ASV spoof challenge results}
\end{table*}

One strategy to spoof an ASV system would be to generate and present speech signals whose features match well those corresponding to genuine, bona fide speech produced by the target speaker (the identity being claimed by the attacker).
In this sense, the characteristics of a speech signal that distinguish different speakers are unlikely to be the same as those that distinguish between bona fide and spoofed speech.  
These assumptions provided the motivation behind the development of CQCCs; instead of using features designed specifically for ASV, a popular choice in the early years of anti-spoofing research, we sought to use features that were fundamentally \emph{different} to those designed for ASV.  
The fundamental assumption was that, if the spoofer is trying to produce speech signals that \emph{replicate} the same features used for ASV, then features optimised for ASV cannot be the best approach to detect spoofed speech signals; features not optimised for ASV would have better potential for spoofing detection since the spoofer was not trying to produce speech signals which  \emph{replicate} these features. 
Although it was discovered later that CQCCs were also useful for other speech tasks including speaker diarization, ASV and utterance verification~\cite{todisco2017constant,kinnunen2016utterance}, they were not \emph{designed} for ASV.  The inspiration for CQCCs came from prior experience of using the constant Q transform (CQT) for music processing tasks~\cite{schorkhuber2014matlab}.

\subsection{From the CQT to CQCCs}
\label{section:CQCC}
The perceptually motivated CQT~\cite{Youngberg78,Brown91} approach to the spectro-temporal analysis of a discrete signal $x(n)$ is defined by:
\begin{equation}
X^{CQ}(k,n)=\sum_{l=n-\left \lfloor N_k/2 \right \rfloor}^{n+\left \lfloor N_k/2 \right \rfloor}x(j)a_{k}^{*}(l-n+N_{k}/2)
\label{eq:CQT}
\end{equation}
where $n$ is the sample index, $k = 1, 2,..., K$ is the frequency bin index, $a_k(n)$ are the basis functions, $*$ is the complex conjugate and $N_k$ is the frame length.
The basis functions $a_k(n)$ are defined by:
\begin{equation}
 a_k(n)=  g_k(n) e^{j 2\pi n {f_{k}}/{f_{s}}},  n \in \mathbb{Z}
\label{eq:atoms}
\end{equation}
where $g_k(n)$ is zero-centred window function, $f_{k}$ is the center of each frequency bin and $f_{s}$ is the sampling rate.
Further details are available in \cite{schorkhuber2014matlab}.
 
The center of each frequency bin $f_{k}$ is defined according to
where $f_{1}$ is the center of the lowest frequency bin and $B$ is the number of bins per octave. $B$ determines the trade-off between spectral and temporal resolutions. 
$N_k \in \mathbb{R}$ in Eqs.~\ref{eq:CQT} and \ref{eq:atoms} is real-valued and inversely proportional to $f_k$. 
As a result, the summation in Eq.~\ref{eq:CQT} is over a number of samples that is dependent upon the frequency.
The spectro-temporal resolution is hence also dependent upon the frequency.
The quality (Q) factor, given by $Q=f_k/\delta f$, where $\delta f$ is the bandwidth, reflects the selectivity of each filter in the filterbank.
For the CQT transform, Q is constant for all frequency bins $k$; filters are logarithmically spaced.

The CQT gives $X^{CQ}(k,n)$ a geometrically spaced spectrum.  
This is resampled using a spline interpolation method to a uniform, linear scale, giving $\bar{X}^{CQ}(l,n)$ which attributes equal weighting to information across the full spectrum~\cite{todisco2016new}.  
CQCCs~\cite{todisco2016new} are obtained from the discrete cosine transformation (DCT) of the logarithm of the squared-magnitude CQT as follows:

\begin{equation}
CQCC(p,n)=
\sum_{l=0}^{L-1}\mathrm{log}\left  |\bar{X}^{CQ}(l,n)  \right |^{2} \mathrm{cos}\left [ \frac{p\left ( l-\frac{1}{2}  \right )\pi}{L} \right ]
\label{eq:cqcc}
\nonumber
\end{equation}
where $p = 0, 1,..., L-1$ and where $l$ is now the linear-scale frequency bin index.
Full details of the CQCC extraction algorithm are available in~\cite{todisco2016new}.
Efficient implementations of the CQT can be found in~\cite{schorkhuber2014matlab} and~\cite{Velasco11}.

\subsection{Differences to the DFT and LFCC}
The principal differences between the CQT and the discrete Fourier transform (DFT) relate to the spectro-temporal resolution.  
Whatever the algorithm, spectral decompositions essentially act as a filterbank. 
For the DFT, the set of filterbank frequencies are linearly distributed and the bandwidth of each filter is constant.  For the DFT, Q is no longer constant, but is linearly distributed. The series of filters is no longer logarithmically spaced, but linearly spaced. This ensures that the DFT exhibits constant spectral resolution.
This is not the case for the CQT. Compared to the DFT and as a direct consequence of the constant Q property, the CQT-derived spectrum has greater frequency resolution at lower frequencies than at higher frequencies.
The CQT-derived spectrogram will then exhibit greater temporal resolution at higher frequencies than at lower frequencies.

\subsection{Performance}
Both CQCC and LFCC front-ends are typically used with simple Gaussian mixture model (GMM) classifiers. The performance obtained with GMM-CQCC and GMM-LFCC systems in terms of EER for the evaluation partition of the ASVspoof 2015 database is illustrated in Table~\ref{Tab:2015 ASV spoof challenge results}, which reproduces results from~\cite{todisco2016new}.
Results are shown independently for each of the 10 different attacks. 
Results for two attacks show marked differences for the CQCC and LFCC front-ends. These are S8, for which LFCCs outperform CQCCs by 93\% relative, and S10 for which CQCCs outperform LFCCs by 87\% relative. While the motivation for our work stems from these two differences for the ASVspoof 2015 database, results reported later in this paper relate to the more recent ASVspoof 2019 database.  

 \section{Sub-band analysis}
The differences in the spectro-temporal resolution of the DFT and CQT lead to one potential explanation for differences in the performance of GMM-LFCC and GMM-CQCC systems.
This relates to differences in the spectral resolution at lower frequencies and temporal resolution at higher frequencies.
Put differently, it might suggest that the artefacts that distinguish spoofed speech from bona fide speech might reside in specific sub-bands, rather than in the full band signal.  
This hypothesis is supported by~\cite{quatieri2006discrete,deller1993discrete,sriskandaraja2016investigation,yang2019significance} which all found this to be the case for synthetic speech.
Accordingly, we set out to determine whether differences at the sub-band level could explain differences in the performance of LFCC and CQCC representations.
Given the stark difference in temporal resolution at high frequencies, our hypothesis has always been that this is where the artefacts are located for attacks such as S10.
The assumption is that these artefacts are then only captured by using a spectro-temporal decomposition with a higher temporal resolution.

\subsection{Experiments}
The set of experiments designed to test this hypothesis consist in an extensive sub-band analysis whereby the GMM-LFCC and GMM-CQCC classifiers are applied to the ASVspoof 2019 logical access database at a sub-band level.
In any single experiment, the entire database is processed with a low-pass and/or high-pass filter.
With both low-pass and high-pass filters, the result is a band-pass filter with cut-in $f_{min}$ and cut-off $f_{max}$.
Corresponding GMM models are retrained each time. 
Cut-in and cut-off frequencies are varied in steps of 400~Hz between 0~Hz and the sampling frequency of 8~kHz. The EER is then determined in the usual manner~\cite{brummer2013bosaris}. 

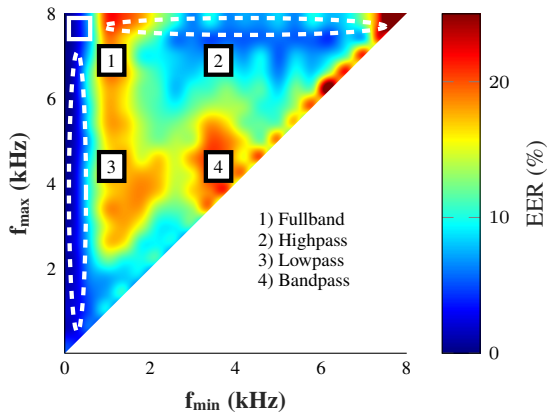
\begin{figure}[!t]
	\centering
	\input{figures/triangle_test.tex}
	\caption{A 2-dimensional heatmap visualisation of sub-band analysis results for an arbitrary spoofing attack.  The horizontal axis depicts the high-pass cut-in frequency $f_{\text{min}}$ whereas the vertical axis depicts the low pass cut-off frequency $f_{\text{max}}$.  The colour bar to the right of the plot depicts the EER obtained for each band-pass filter configuration (pair of $f_{\text{min}}$ and $f_{\text{max}}$).}

	\label{Figure. triangle explaination}
\end{figure}

\subsection{Heatmap visualisation}
Results are visualised in the form of a 2-dimensional heatmap, an example of which is illustrated in Fig.~\ref{Figure. triangle explaination}.
The horizontal vertical axes of the heatmap signify the cut-in frequency $f_{\text{min}}$ and cut-off frequency $f_{\text{max}}$ of the high-pass and low-pass filter respectively.
For band-pass filters,  $f_{\text{min}}<f_{\text{max}}$, hence the triangular form.  
The EER corresponding to each band-pass filter configuration is indicated with the colour bar to the right of Fig.~\ref{Figure. triangle explaination}, with blue colours indicating low EERs and red colours indicating higher EERs.

The left-most column of the heatmap shows the EER for a decreasingly aggressive low-pass filter, moving from bottom-to-top. 
The top-most row of the heatmap shows the EER for a increasingly aggressive high-pass filter moving from left-to-right.
Everywhere else corresponds to a band-pass filter, with the full-band configuration being at the top-left.
EERs along the diagonal of Fig.~\ref{Figure. triangle explaination} illustrate the benefit of spectrum information at the sub-band level.

The heatmap serves as a crude means to evaluate which sub-bands are the least and most informative in distinguishing between spoofed and bona fide speech. 
For the arbitrary example illustrated in Fig.~\ref{Figure. triangle explaination}, EERs along the diagonal suggest that information at lower frequencies is discriminative, whereas use of that at higher frequencies is not sufficiently discriminative to be used alone. 
EERs in the left-most column show that the most discriminative information of all is contained at low frequencies.
So long as this information is used, then the EER is low. 
EERs in the top-most row show that as soon as low frequency information is discarded, then the EER increases.  
Information between~1 and 3~kHz and above 7~kHz is non-discriminative.
That between~3 and 7~kHz is less discriminative, with reasonable EERs being obtained only when the information in a sufficient number of sub-bands is combined.

\section{Experiments with the \\ASVspoof 2019 database}
  
Described here is our use of the ASVspoof 2019 database, the specific configurations of GMM-LFCC and GMM-CQCC spoofing countermeasures, the subset of spoofing attacks with which this work was performed, and the results of sub-band analysis.

\subsection{Database and protocols}

The ASVspoof 2019 database consist of two different protocols based on logical access (LA) and physical access (PA) use case scenarios~\cite{todisco2019asvspoof}.
The LA scenario relates to spoofing attacks performed with voice conversion and speech synthesis algorithms whereas the PA scenario involves exclusively replay spoofing attacks.
Since the work reported in this paper stems from observations made for spoofing attacks performed with voice conversion and synthetic speech algorithms, our experiments were performed with the LA dataset only.

All experiments were performed with the standard ASVspoof 2019 protocols which consist of independent train, development and, evaluation partitions. 
Spoofed speech in each partition is generated using a set of voice conversion and speech synthesis algorithms~\cite{wang2019asvspoof}. 
There are 19 different spoofing attacks.
Attacks in the training and development set were created with a set of 6 algorithms (A01-A06), whereas  those in the evaluation set were created with a set of 13 algorithms (A07-A19). 
All experiments reported in this paper were performed with the evaluation set only.

\begin{table*}[!t]
	\centering
	\caption{EERs (\%) for the ASVspoof 2019 logical access database, evaluation partition. Results for both baseline systems, GMM-CQCC (linear scale), GMM-LFCC and GMM-CQCC (geometric scale).  Results for GMM-CQCC (linear) and GMM-LFCC reproduced from~\cite{todisco2019asvspoof}.  Results for six attacks (highlighted) show stark differences in performance for each system.}
	\setlength\tabcolsep{4.2pt}
	\begin{tabular}{ *{14}{c}}
		\hline
	    System	&\textbf{A07} &A08&A09&A10&A11&A12&\textbf{A13}&\textbf{A14}&A15&\textbf{A16}&\textbf{A17}&A18 &\textbf{A19}	\\ 
		\hline\hline
		GMM-CQCC (linear)&\color{blue}{\textbf{0.00}}&0.04&0.14&15.16&0.08&4.74&
		\color{purple}{\textbf{26.15}}&\color{purple}{\textbf{10.85}}&1.26&\color{blue}{\textbf{0.00}}&\textbf{19.62}&3.81&\color{blue}{\textbf{0.04}} \\		
		\hline 
		GMM-LFCC&\color{purple}{\textbf{12.86}}&0.37&0.00&18.97&0.12&4.92&\textbf{9.57}&\textbf{1.22}&2.22&\color{purple}{\textbf{6.31}}&\color{blue}{\textbf{7.71}}&3.58&\textbf{13.94}\\
		\hline 
		GMM-CQCC (geometric)&\textbf{3.39}&0.34&0.46&6.86&4.62&3.58&
		\color{blue}{\textbf{4.23}}&\color{blue}{\textbf{0.67}}&1.52&\textbf{4.00}&\color{purple}{\textbf{25.04}}&19.63&\color{purple}{\textbf{29.46}}\\
		\hline
	\end{tabular}
	\label{Tab:2019 ASV spoof results}
\end{table*}

\begin{table*}[!t]
\centering
\caption{A summary of the six spoofing attack algorithms (voice conversion and speech synthesis) from the ASVspoof 2019 logical access database used in this work. * indicates neural networks. Full details for each algorithm can be found in~\cite{wang2019asvspoof}.}
\resizebox{\textwidth}{!}{%
\begin{tabular}{|c|c|c|c|c|c|c|c|}
\hline
 Attack & Input & Input processor & Duration & Conversion & Speaker represent. & Outputs & Waveform generator \\ \hline
A07~\cite{wavecyclegan,wu2016merlin} 
& Text & NLP & RNN* & RNN* & One hot embed. & MCC, F0, BA & WORLD \\ \hline
A13~\cite{li2015generative,kobayashi2018intra} 
& Speech (TTS) & WORLD & DTW & Moment matching* & - & MCC & Waveform filtering \\ \hline
A14~\cite{ljliu2018wav,Kawahara1999Restructuring} 
& Speech (TTS) & ASR* & - & RNN* & - & MCC, F0, BAP & STRAIGHT \\ \hline
A16~\cite{McAuliffe/etal:2017:IS} & Text & NLP & - & CART & - & MFCC, F0 & Waveform concat. \\ \hline
A17~\cite{2019arXiv190711898H,kobayashi2018intra}
& Speech (human) & WORLD & - & VAE* & One hot embed. & MCC, F0 & Waveform filtering \\ \hline
A19 & Speech (human) & LPCC/MFCC & - & GMM-UBM & - & LPC & Spectral filtering + OLA \\ \hline
\end{tabular}%
}
\label{tab:asvspoof_2019_attack_details}
\end{table*}

\subsection{Countermeasures}
 
Experiments were performed using the two spoofing countermeasures that were distributed with the ASVspoof 2019 challenge database.  
They are the GMM-LFCC and GMM-CQCC baseline systems\footnote{\url{https://www.asvspoof.org/asvspoof2019/asvspoof2019_evaluation_plan.pdf}}~\cite{todisco2019asvspoof}.
The front-end representation used by the GMM-LFCC system comprises 20 static, velocity ($\Delta$) and acceleration ($\Delta \Delta$) LFCC coefficients, thereby giving 60-dimension feature vectors.
The 90-dimension GMM-CQCC representation comprises 29 static with zero energy coefficient, $\Delta$, $\Delta\Delta$ coefficients.
The back-end is common to both systems and consists of a traditional GMM classifier with two models, one for bona fide speech and one for spoofed speech, both with 512 models.
Scores are conventional log-likelihood ratios.
Neither countermeasure system was re-optimised; this work is not concerned with optimisation.  
Results reported here are those obtained with the original ASVspoof 2019 baseline countermeasures as reported in~\cite{todisco2019asvspoof}.

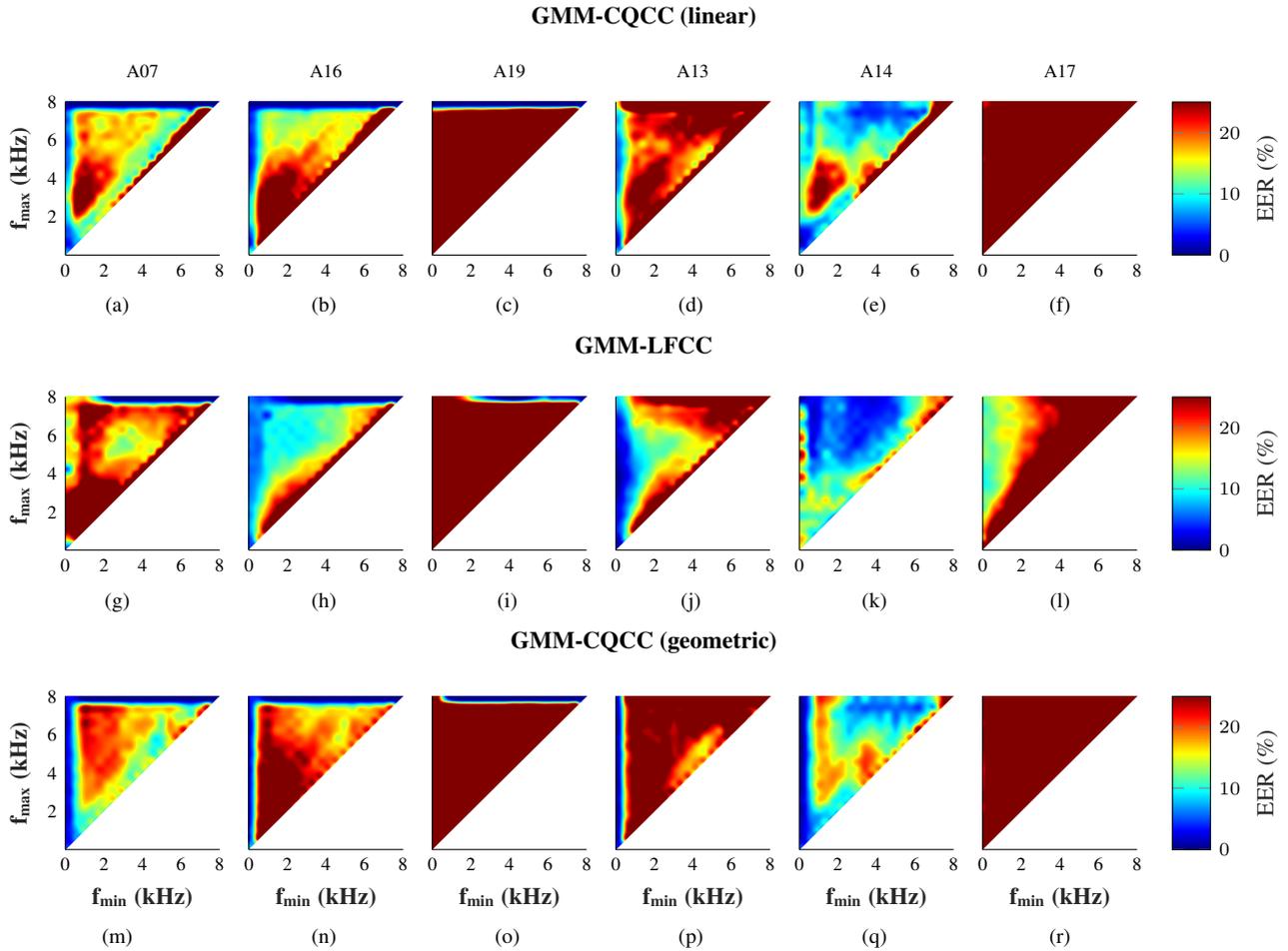
\begin{figure*}[!t]
     \centering
     {\textbf{GMM-CQCC (linear)}}
     \subfloat[\label{fig:triangles:A07_lin_cqt}]{\input{figures/A07_linear_CQT_clean}}
     \subfloat[\label{fig:triangles:A16_lin_cqt}]{\input{figures/A16_linear_CQT_clean}}
     \subfloat[\label{fig:triangles:A19_lin_cqt}]{\input{figures/A19_linear_CQT_clean}}
     \subfloat[\label{fig:triangles:A13_lin_cqt}]{\input{figures/A13_linear_CQT_clean}}
     \subfloat[\label{fig:triangles:A14_lin_cqt}]{\input{figures/A14_linear_CQT_clean}}
     \subfloat[\label{fig:triangles:A17_lin_cqt}\protect\hphantom{MAKESPACE}]{\input{figures/A17_linear_CQT_clean}}
     \\
     \vspace{0.2cm}
     {\textbf{GMM-LFCC }}
     \subfloat[\label{fig:triangles:a}]{\input{figures/A07_linear_FFT_clean}}
     \subfloat[\label{fig:triangles:a}]{\input{figures/A16_linear_FFT_clean}}
     \subfloat[\label{fig:triangles:a}]{\input{figures/A19_linear_FFT_clean}}
     \subfloat[\label{fig:triangles:a}]{\input{figures/A13_linear_FFT_clean}}
     \subfloat[\label{fig:triangles:a}]{\input{figures/A14_linear_FFT_clean}}
     \subfloat[\label{fig:triangles:a}\protect\hphantom{MAKESPACE}]{\input{figures/A17_linear_FFT_clean}}
     \\
     \vspace{0.2cm}
     {\textbf{GMM-CQCC (geometric)}}
     \subfloat[\label{fig:triangles:a}]{\input{figures/A07_geometric_CQT_clean.tex}}
     \subfloat[\label{fig:triangles:a}]{\input{figures/A16_geometric_CQT_clean.tex}}
     \subfloat[\label{fig:triangles:a}]{\input{figures/A19_geometric_CQT_clean.tex}}
     \subfloat[\label{fig:triangles:a}]{\input{figures/A13_geometric_CQT_clean.tex}}
     \subfloat[\label{fig:triangles:a}]{\input{figures/A14_geometric_CQT_clean.tex}}
     \subfloat[\label{fig:triangles:a}\protect\hphantom{MAKESPACE}]{\input{figures/A17_geometric_CQT_clean.tex}}
     \caption{2-D heatmap visualisations of sub-band analysis results for the six different spoof attacks.  The first row (a-f) shows results for the GMM-CQCC (linear) system.  The second row (g-l) shows corresponding results for the GMM-LFCC system.  The third row (m-r) shows results for the GMM-CQCC (geometric) system.}
     \label{fig:triangles}
 \end{figure*}

\subsection{Metric and performance}

Even though it is not the primary metric for the ASVspoof 2019 database, performance is assessed in terms of the EER which is determined according to a convex hull approach \cite{brummer2013bosaris}.
While the community has moved towards use of the tandem detection cost function (t-DCF) metric~\cite{kinnunen2018t}, the primary metric for the ASVspoof 2019 database, the work reported in this paper is concerned exclusively with countermeasure performance rather than the impact of spoofing and countermeasures upon ASV performance.
The latter is a related, but different issue.
Use of the EER is then appropriate and sufficient for this purposes of the work presented in this paper.

EER results for the two countermeasures and the evaluation partition of the ASVspoof 2019 LA database are illustrated in Table~\ref{Tab:2019 ASV spoof results}.
Just as is the case for the 2015 database, results for the 2019 database show substantial variations in performance for the GMM-LFCC and GMM-CQCC countermeasures. 
For attacks A07, A16 and A19, the GMM-CQCC system outperforms GMM-LFCC system whereas, for attacks A13, A14 and A17, it is the GMM-LFCC system that performs best, albeit still with relatively high error rates.

Subsequent experiments were performed separately for this subset of 6 specific spoofing attacks, all being examples of where one front-end representation leads to substantially better results than the other. Brief details of the specific algorithms used in creating each attack are illustrated in Table~\ref{tab:asvspoof_2019_attack_details}.  
There are four voice conversion algorithms (A13, A14, A17 and A19) and two speech synthesis algorithms (A07 and A16), even if the input to two of the voice conversion algorithms is also synthetic speech (A13 and A14).
Further details of each algorithm are available in~\cite{wang2019asvspoof}.

\subsection{Sub-band analyses}

We sought to explain differences in the performance of GMM-LFCC and GMM-CQCC systems using what we could learn from sub-band analysis, results for which are shown in Fig.~\ref{fig:triangles}.
Heatmaps are shown for each of the 6 spoofing attacks: A07, A16, A19 for which the GMM-CQCC system performs best; A13, A14 and A17 for which the GMM-LFCC system performs best.
Results for the GMM-CQCC and GMM-LFCC systems are illustrated in rows one and two respectively. In each case, the EER for the baseline system is denoted by the top-left-most point in each 2D heatmap, i.e.~the full-band case.

Figs.~\ref{fig:triangles} (a)-(c) show that the baseline GMM-CQCC system (top-left-most point) gives low EERs whereas Figs.~\ref{fig:triangles} (g)-(i) show that the GMM-LFCC system gives consistently worse results.
Figs.~\ref{fig:triangles} (d)-(f) and (j)-(l) show the opposite, even if performance for A17 (l) is still poor for both the baseline systems.
Upon inspection of heatmaps for both LFCC and CQCC front-ends, we see that the critical information for the detection of all three attacks lies at high frequencies.
While it is not visible in the plots at this scale, the discriminative information lies above 7.6~kHz; near-to-zero EERs can be obtained using information between~7.6 and 8~kHz only, using either front-end.   

The situation for attacks A13, A14 and A17 is a little different.
The diagonals of the heatmaps in Figs.~\ref{fig:triangles} (d), (e), (j) and (k) all suggest that the most informative information is at the lower frequencies.
The left-most columns of Figs.~\ref{fig:triangles} (d), (e) and (j) suggest that the most critical information is located at the very lowest sub-bands.
For A17, neither front-end performs especially well.
The CQCC front-end fails completely, whereas the LFCC front-end succeeds in capturing information across the majority of the full band.

 \begin{figure*}[!t]
	\centering
	\includegraphics{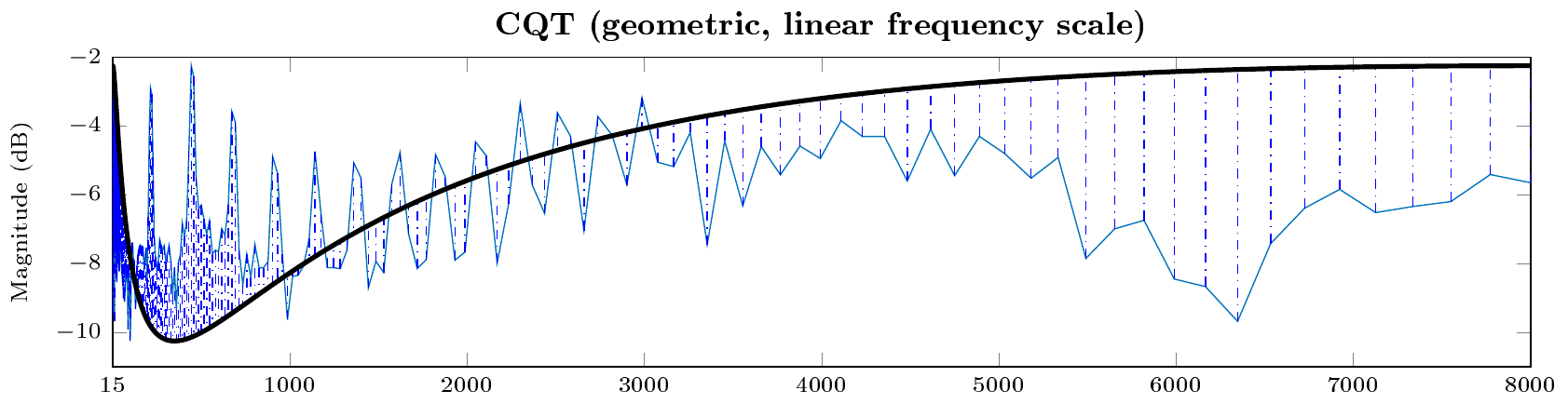}\\
	\includegraphics[width=\textwidth,trim={0cm 0cm 0cm 7.5cm},clip]{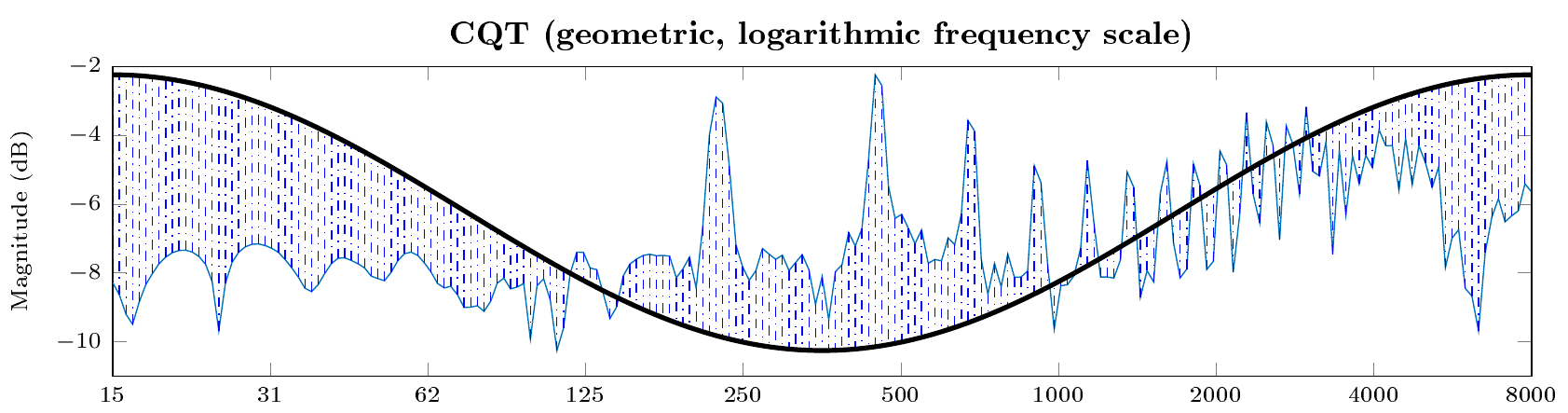}\\
	\includegraphics{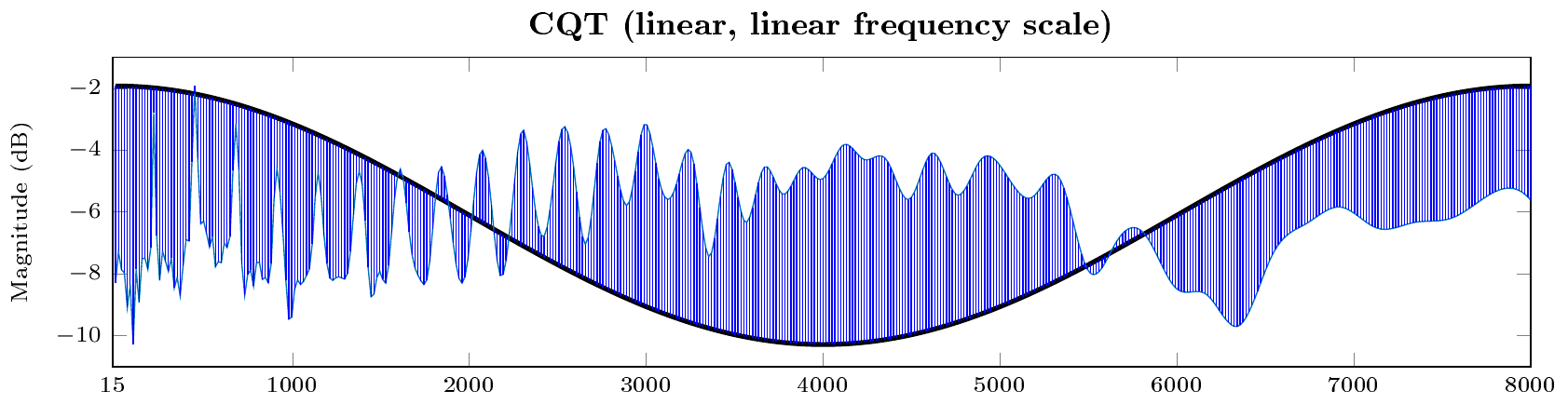}\\
	\includegraphics{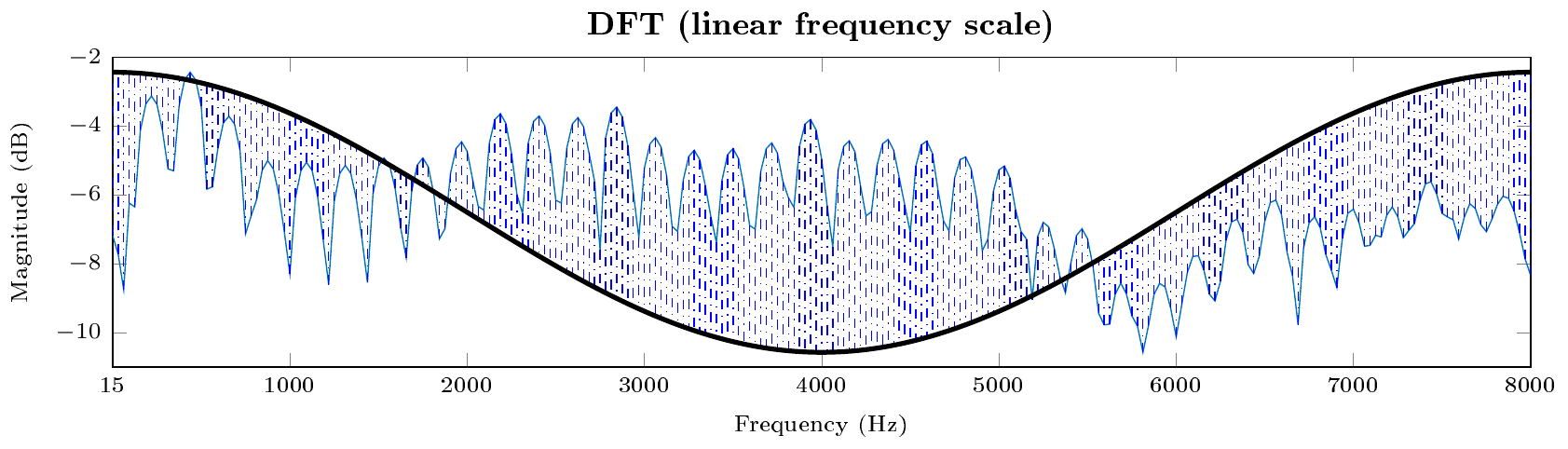}\\

	\caption{Illustrations of spectra for an arbitrary speech frame derived using the CQT and DFT (blue) and the second basis function of the discrete cosine transform used in cepstral analysis (black). The top plot shows the CQT-derived spectrum without resampling, hence the logarithmic sampling (compression of vertical blue lines) in the linear frequency domain. The second plot shows the same spectrum but with a warped, logarithmic frequency scale.  The third plot shows the CQT-derived spectrum after resampling to a uniform, linear frequency scale (Section 2; thus more vertical lines).  The forth plot shows the corresponding DFT-derived spectrum.}
	\label{Figure.resample}
\end{figure*}

\section{Spectro-temporal resolution} 

Results presented above dispel somewhat the hypothesis that CQCCs produces better results than LFCCs for some attacks on account of the higher temporal resolution at high frequencies. 
If this were true, reliable performance would not have been obtained with the LFCC front-end using high frequency sub-bands alone.
The explanation lies elsewhere.

The same results show that the discriminative information simply lies in the highest sub-bands; with appropriate band-pass filtering, both front-ends perform well for A07, A16 and A19 attacks.
It is more than this, though, since such a straightforward explanation does not account for why the original, \emph{full-band} CQCC front-end performs well, i.e.~\emph{without} band-pass filtering.
The explanation for this observation requires us to revisit the issue of spectro-temporal resolution.

Fig.~\ref{Figure.resample} illustrates a set of DFT and CQT-derived spectra for an arbitrary speech frame. Each plot also shows the second basis function of the DCT used in cepstral analysis.
The vertical bars serve to illustrate the \emph{spectral} resolution across the spectrum and give some indication of the \emph{temporal} resolution, which is inversely proportional to the spectral resolution. 

The top plot in Fig.~\ref{Figure.resample} shows the CQT-derived spectrum (without re-sampling)~\cite{brown1999computer}.  
It shows clearly that the spectral resolution is higher at lower frequencies than at higher frequencies.  
The DCT basis function in this plot is \emph{warped} to the same, non-linear frequency scale. 
The second plot shows exactly the same data, now plotted with a logarithmic frequency scale, hence the regularity of the vertical bars and DCT basis function.
These plots show how, without resampling, the DCT will attribute greater \emph{weight} to the low frequency components than to high frequency components.  
The third plot shows the CQT-derived spectrum after resampling.
The the vertical bars show how resampling acts to linearise the rate of sampling in the frequency domain.  
Note the difference in frequency scales for the second and third plots.
The higher spectral resolution (and hence lower temporal resolution) for lower frequencies 
is still clearly apparent; the spectrum is much smoother at higher frequencies.

A comparison of the cosine basis functions to the CQT-derived spectra in the second and third plots shows that resampling acts to dilute (weaken) information at lower frequencies but to distil (emphasise) information at higher frequencies. Hence, spoofing artefacts at high frequencies are emphasised by the CQCC front-end.

The fourth plot of Fig.~\ref{Figure.resample} shows the DFT-derived spectrum and the linear sampling in the frequency domain. 
Here, the DCT acts to weight all frequency components uniformly.
Spoofing artefacts at high frequencies are not emphasised; reliable performance is then obtained only by band-pass filtering.

The explanation for why the LFCC front-end outperforms the CQCC front-end for attacks A13, A14 and A17 is now straightforward.
The artefacts for these attacks are at lower and mid-range frequencies where the CQT acts to dilute information, hency why the GMM-CQCC system performs poorly for these attacks.
Since the DFT gives uniform weighting to information at all frequencies, the GMM-LFCC system gives better, though still somewhat poor performance.
This is because information at low frequencies is not \emph{emphasised}.

\section{Validation}

We designed one final experiment to validate our findings.
The CQT-derived spectra in the top two plots of Fig.~\ref{Figure.resample} show dense sampling of the spectrum at lower frequencies but sparse sampling at higher frequencies.
Hence, without resampling, the application of cepstral analysis should result in the emphasis of information situated at lower frequencies.
In this case, a CQCC front-end \emph{without} linear resampling should produce better detection performance in the case of attacks A13 and A14.  
Performance for attack A17 might not be improved, since the artefacts appear not to be localised at low frequencies.

We repeated the sub-band analysis described in Section~3 using the geometrically-scaled CQCC front-end.
Results are illustrated in Figs.~\ref{fig:triangles} (m)-(r) for the same set of 6 attacks for which results are described in Section~4.
The comparison of the right-most columns in the heatmaps of Figs.~\ref{fig:triangles} (d) and (p) and those in (e) and (q) clearly show that lower EERs are obtained using the geometrically-scaled rather than the linearly-scaled CQCC front-end.
Low EERs are even obtained when the front-end is applied without any band-pass filtering.
This is because use of the original CQT geometric frequency scale results in the emphasis of information at lower frequencies where the artefacts are located.
As expected, attack A17 remains troublesome.

EERs for the geometrically-scaled CQCC front-end are presented in the last row of Table~\ref{Tab:2019 ASV spoof results}.
For some attacks, the EER for the geometrically-scaled CQCC front-end is slightly higher.
For some others, it is slightly lower.
Once again, performance for A17 is poor, and even worse than for both the linearly-scaled CQCC and LFCC front-ends.
Those for A13 and A14 clearly confirm our findings, with substantially lower EERs than those obtained with both the linearly-scaled CQCC and LFCC front-ends.
Use of the original geometric scale of the CQT clearly results in the emphasis of information at lower frequencies where the artefacts are localised.

\section{Discussion}
This paper reports an explainability study of constant Q cepstral coefficients (CQCCs), a spoofing countermeasure front-end for automatic speaker verification~\cite{todisco2016new,todisco2017constant}.
The work aims to explain why the CQCC front-end works so reliably in detecting some forms of spoofing attack, but performs poorly in detecting others.
It confirms that different spoofing attacks exhibit artefacts at different frequencies, artefacts that are better captured with specific front-ends.
The standard constant Q transform (CQT)~\cite{Youngberg78,Brown91} exhibits a dense sampling of the spectrum at lower frequencies and a sparse sampling at higher frequencies. Hence, geometrically-sampled CQCCs perform well in detecting spoofing artefacts when they are located at low frequencies.
Linear sampling shifts the emphasis to higher frequencies so that the detection of spoofing artefacts at high frequencies are emphasised and captured reliably.

Taken together, the results presented in this paper show that no single CQCC front-end configuration can perform well for all spoofing attacks; different spoofing attacks produce artefacts at different parts of the spectrum and these can only be detected reliably when the front-end emphasises information in the relevant frequency bands.
This finding may explain why classifier fusion has proven to be so important to generalisation, i.e.~reliable performance in the face of varying spoofing attacks.
Given that spoofing artefacts can be highly localised in the spectrum, the same finding suggests that we may need to rethink the use of cepstral analysis, which acts to smooth information located across the full spectrum.
Approaches to detect localised artefacts need further investigation for multiple applications (the EER represents just one operating point).
We may then need to rethink the approach to classifier fusion too.

One natural extension of the work would be to look deeper into the artefacts themselves.
Preliminary investigations show that those at high frequencies appear to be nothing more than band-limiting effects introduced by the spoofing algorithms, or the data used in their training. The source of those at lower frequencies are less clear. 
A clearer understanding of the artefacts at the signal level, rather than just the feature or spectrum level is certainly needed.
This could help us to design better spoofing countermeasures, even if the same understanding  might also help the fraudsters to design better spoofing attacks. 

Lastly, while it was not the objective of this work to investigate the generalisation capabilities of the CQCC front-end, it is clear that none of the countermeasures for which results are reported in this paper is successful in detecting \emph{all} of the spoofing attacks used in the creation of the ASVspoof 2019 database.
Attacks A10, A12, A13, A17 and A18 are all examples where \emph{neither} configuration of the CQCC front-end, \emph{nor} the LFCC front-end are especially effective.
The work presented in this paper focused on countermeasure performance exclusively.
It did not consider the effect of spoofing and countermeasures upon automatic speaker verification (ASV) performance.
Some attacks, e.g.~A17, are known not to be especially effective in fooling ASV, and in this sense they pose no threat; that countermeasures fail to detect such attacks is hence of no serious concern.
Other attacks, though, are effective in manipulating ASV and are also difficult to detect reliably.
Other work reported by the community has shown far better results than those reported in this paper~\cite{chettri2019ensemble, alluri2019iiit,yang2019sjtu,lavrentyeva2019stc}. 
Still, even the very best approaches can be vulnerable to certain, specific attacks.
The \emph{silver bullet} remains elusive for the time being.

\section{Acknowledgements}
{\small
The work was partially supported by the Voice Personae and RESPECT projects, both funded by the French Agence Nationale de la Recherche~(ANR).
}

\balance
\bibliographystyle{IEEEbib}

\balance
\end{document}

%% file: preambles.tex
\PassOptionsToPackage{hyphens}{url}
\usepackage[bookmarks=false,hidelinks]{hyperref}

\usepackage[caption=false,font=footnotesize,subrefformat=parens,labelformat=parens]{subfig}
\usepackage{tikz}
\usepackage{pgfplots}
\usepackage{grffile}
\usepackage{float}
\pgfplotsset{compat=newest}
\makeatletter
\g@addto@macro\@floatboxreset\centering
\makeatother
\def\addlegendimage{\csname pgfplots@addlegendimage\endcsname}
\usepackage{graphicx}
\usepackage{cmap}
\usetikzlibrary{shapes.misc}
\usetikzlibrary{mindmap}
\usetikzlibrary{plotmarks}
\usetikzlibrary{arrows.meta}
\usepgfplotslibrary{patchplots}

\tikzset{
    font=\scriptsize
}

\newcommand{\setvariable}[2]{
	\let#1\relax
	\newcommand{#1}{#2}
}

\definecolor{darkgreen}{rgb}{0.2,0.7,0.3}
\definecolor{darkorange}{rgb}{1,0.55,0}

\pgfplotsset{
    mated/.style={ybar interval, fill opacity=0.6, area legend, 
        fill=darkgreen, draw=white},
    nonmated/.style={ybar interval, fill opacity=0.6, area legend, 
        fill=white!30!red, draw=white},
    histAxis/.style={
        width=\figwidth,
        height=\figheight,
        scale only axis,
        ymin=0,ymax=1,
        ylabel={Rel. freq.},
        xmajorgrids,ymajorgrids,},
    rawHistAxis/.style={histAxis,xlabel={Scores},xmin=-20,xmax=40},
    llrHistAxis/.style={histAxis,xlabel={Ideal LLRs},xmin=-10,xmax=10},
}

\usepackage{eurosym}

\setvariable{\colorScalePurple}{PiYG-11-1}
\setvariable{\colorScaleOrange}{Oranges-9-4}
\setvariable{\colorScaleBlue}{Spectral-5-5}
\setvariable{\colorScaleRed}{hda}
\setvariable{\colorScaleGreen}{Spectral-5-4!80!black}

%% file: figures/triangle_test.tex
%
%
\begin{tikzpicture}
\centering
\begin{axis}[%
width=4.5cm,
height=4.5cm,
at={(0.973in,0.502in)},
scale only axis,
point meta min=0,
point meta max=25,
axis on top,
xmin=0.5,
xmax=2000.5,
xtick={0.5,500,1000,1500,2000},
xticklabels={{0},{2},{4},{6},{8}},
xlabel style={font=\bfseries\color{white!15!black}},
xlabel={$\text{f}_{\text{min}}$} {\text{(kHz)}},
y dir=reverse,
ymin=0.5,
ymax=2000.5,
ytick={0.5,500,1000,1500},
yticklabels={{8},{6},{4},{2}},
ylabel style={font=\bfseries\color{white!15!black}},
ylabel={$\text{f}_{\text{max}}$} {\text{(kHz)}},
axis background/.style={fill=white},
axis x line*=bottom,
axis y line*=left,
xtick style={draw=none},
ytick style={draw=none},
legend style={legend cell align=left, align=left, draw=white!15!black},
colormap/jet,
colorbar,
colorbar style={ylabel style={font=\color{white!15!black}}, ylabel={EER (\%)}}
]
\addplot [forget plot] graphics [xmin=0.5, xmax=2000.5, ymin=0.5, ymax=2000.5] {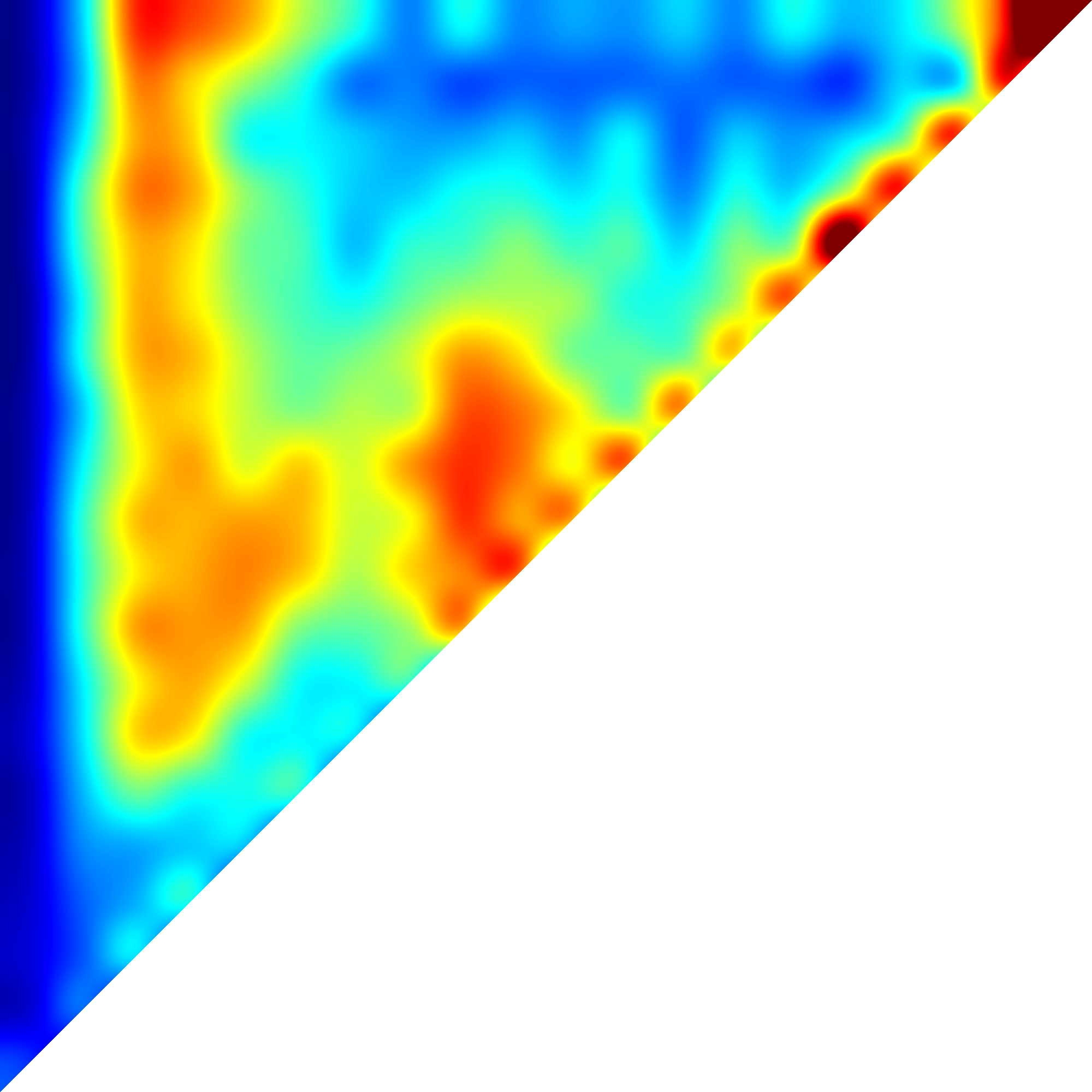};
\draw[white,dashed,ultra thick] (axis cs:75,1050) ellipse[x radius=50, y radius=825];
\draw[white,dashed,ultra thick]  (axis cs:1050,75) ellipse[y radius=50, x radius=825];
\draw[white, ultra thick] (axis cs:25,25)  rectangle  ++(125,125);
\node[fill=white,draw=black,ultra thick] at (axis cs:275,275) {1};
\node[fill=white,draw=black,ultra thick] at (axis cs:900,275) {2};
\node[fill=white,draw=black,ultra thick] at (axis cs:275,900) {3};
\node[fill=white,draw=black,ultra thick] at (axis cs:900,900) {4};
\node[align=left] at (axis cs: 1400,1400) {1) Fullband \\  2) Highpass \\ 3) Lowpass \\ 4) Bandpass};
\end{axis}
\end{tikzpicture}%

%% file: figures/A07_linear_CQT_clean.tex
%
\begin{tikzpicture}
\begin{axis}[%
width=.8in,
height=.8in,
scale only axis,
point meta min=0,
point meta max=25,
axis on top,
xmin=0.5,
xmax=2000.5,
xtick={0.5,500,1000,1500,2000},
xticklabels={{0},{2},{4},{6},{8}},
y dir=reverse,
ymin=0.5,
ymax=2000.5,
ytick={0.5,500,1000,1500},
yticklabels={{8},{6},{4},{2}},
ylabel style={font=\bfseries\color{white!15!black}},
ylabel={$\text{f}_{\text{max}}$} {\text{(kHz)}},
axis background/.style={fill=white},
title={A07},
title style={align=center},
axis x line*=bottom,
axis y line*=left,
xtick style={draw=none},
ytick style={draw=none},
]
\addplot [forget plot] graphics [xmin=0.5, xmax=2000.5, ymin=0.5, ymax=2000.5] {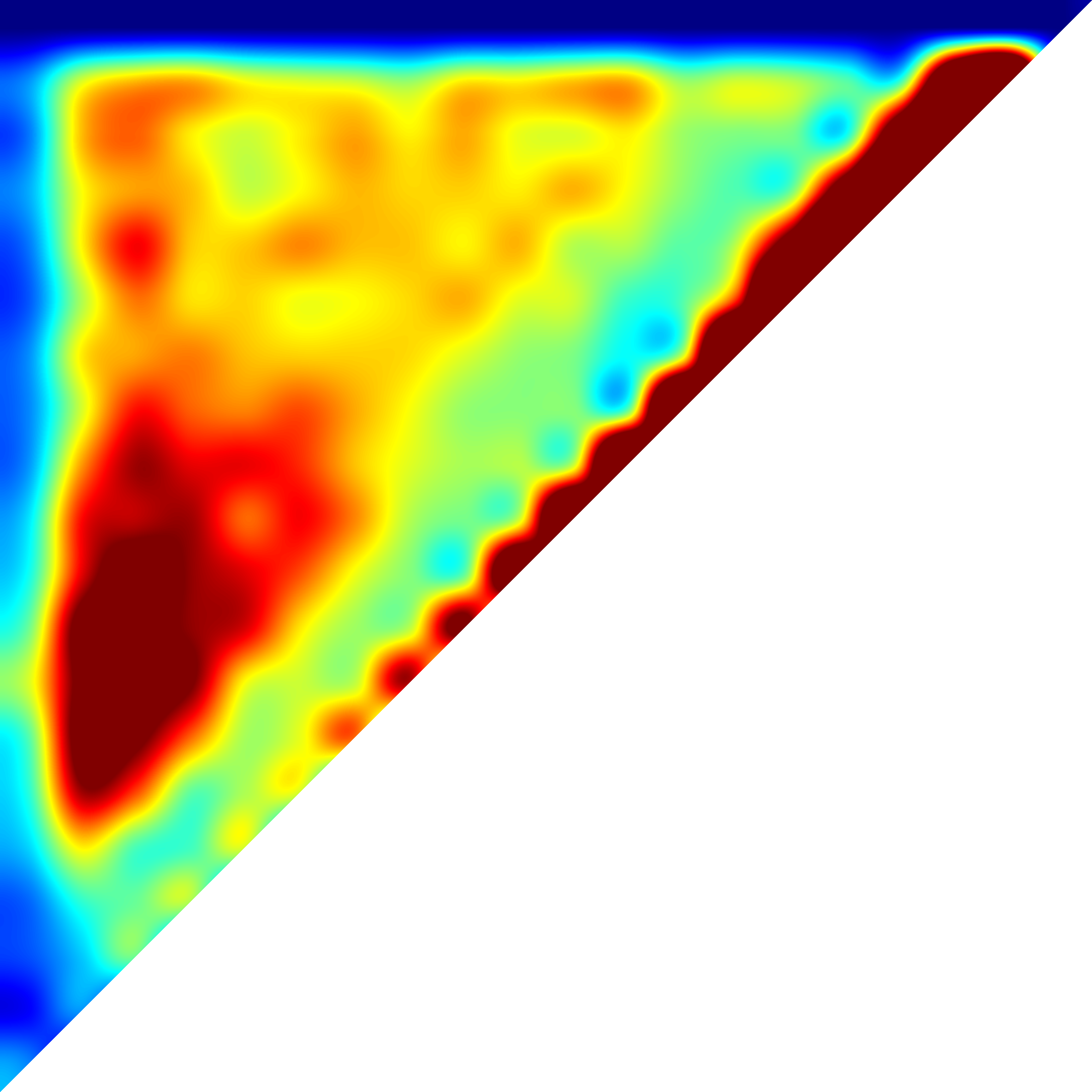};
\end{axis}
\end{tikzpicture}%

%% file: figures/A16_linear_CQT_clean.tex
%
%
\begin{tikzpicture}
\begin{axis}[%
width=.8in,
height=.8in,
at={(1.346in,0.502in)},
scale only axis,
point meta min=0,
point meta max=25,
axis on top,
xmin=0.5,
xmax=2000.5,
xtick={0.5,500,1000,1500,2000},
xticklabels={{0},{2},{4},{6},{8}},
ymin=0.5,
ymax=2000.5,
ytick={},
yticklabels={},
title={A16},
title style={align=center},
axis x line*=bottom,
axis y line*=left,
xtick style={draw=none},
ytick style={draw=none},
legend style={legend cell align=left, align=left, draw=white!15!black}
]
\addplot [forget plot] graphics [xmin=0.5, xmax=2000.5, ymin=0.5, ymax=2000.5] {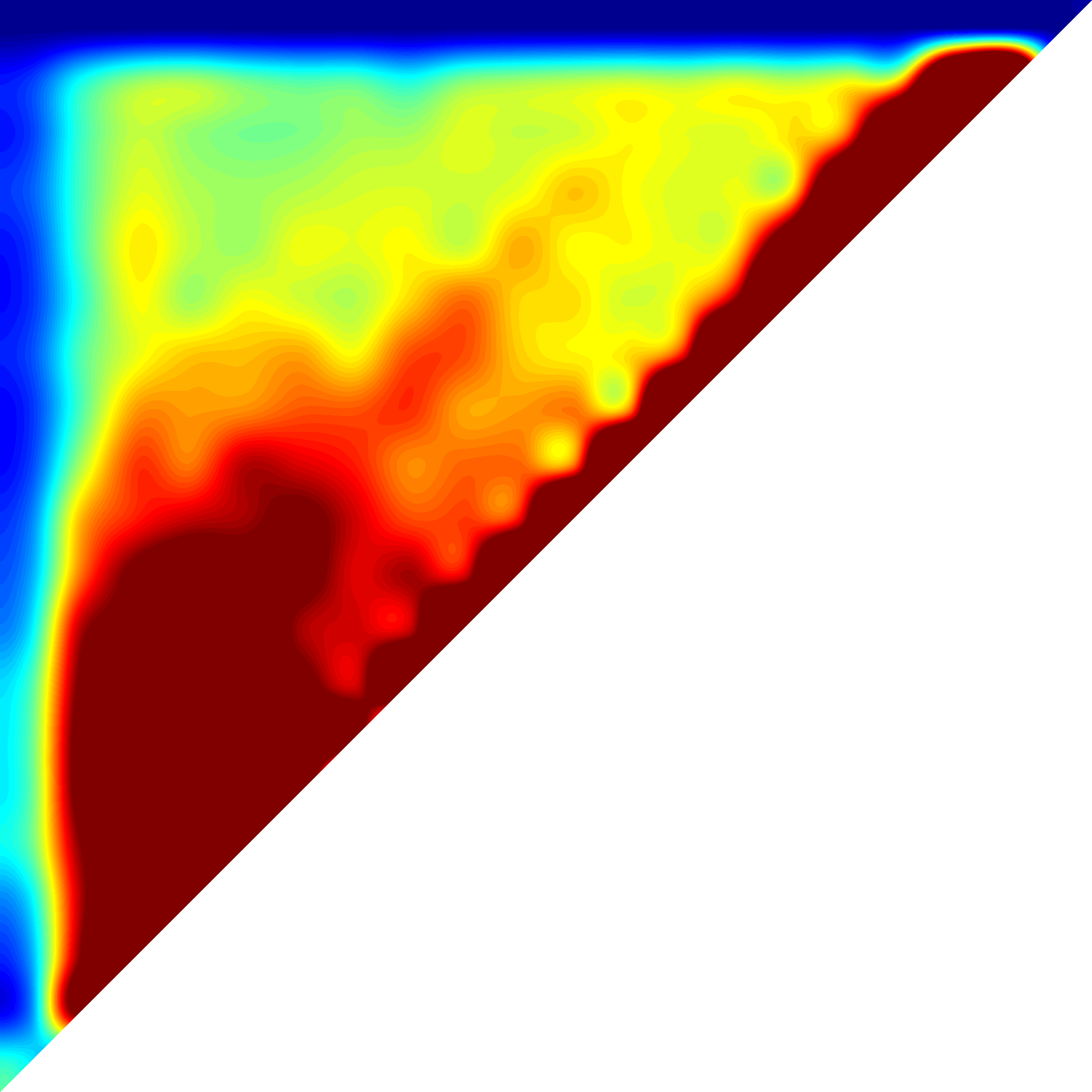};

\end{axis}
\end{tikzpicture}%

%% file: figures/A19_linear_CQT_clean.tex
%
%
\begin{tikzpicture}

\begin{axis}[%
width=.8in,
height=.8in,
at={(1.346in,0.502in)},
scale only axis,
point meta min=0,
point meta max=25,
axis on top,
xmin=0.5,
xmax=2000.5,
xtick={0.5,500,1000,1500,2000},
xticklabels={{0},{2},{4},{6},{8}},
y dir=reverse,
ymin=0.5,
ymax=2000.5,
ytick={},
yticklabels={},
axis background/.style={fill=white},
title={A19},
title style={align=center},
axis x line*=bottom,
axis y line*=left,
xtick style={draw=none},
ytick style={draw=none},
legend style={legend cell align=left, align=left, draw=white!15!black}
]
\addplot [forget plot] graphics [xmin=0.5, xmax=2000.5, ymin=0.5, ymax=2000.5] {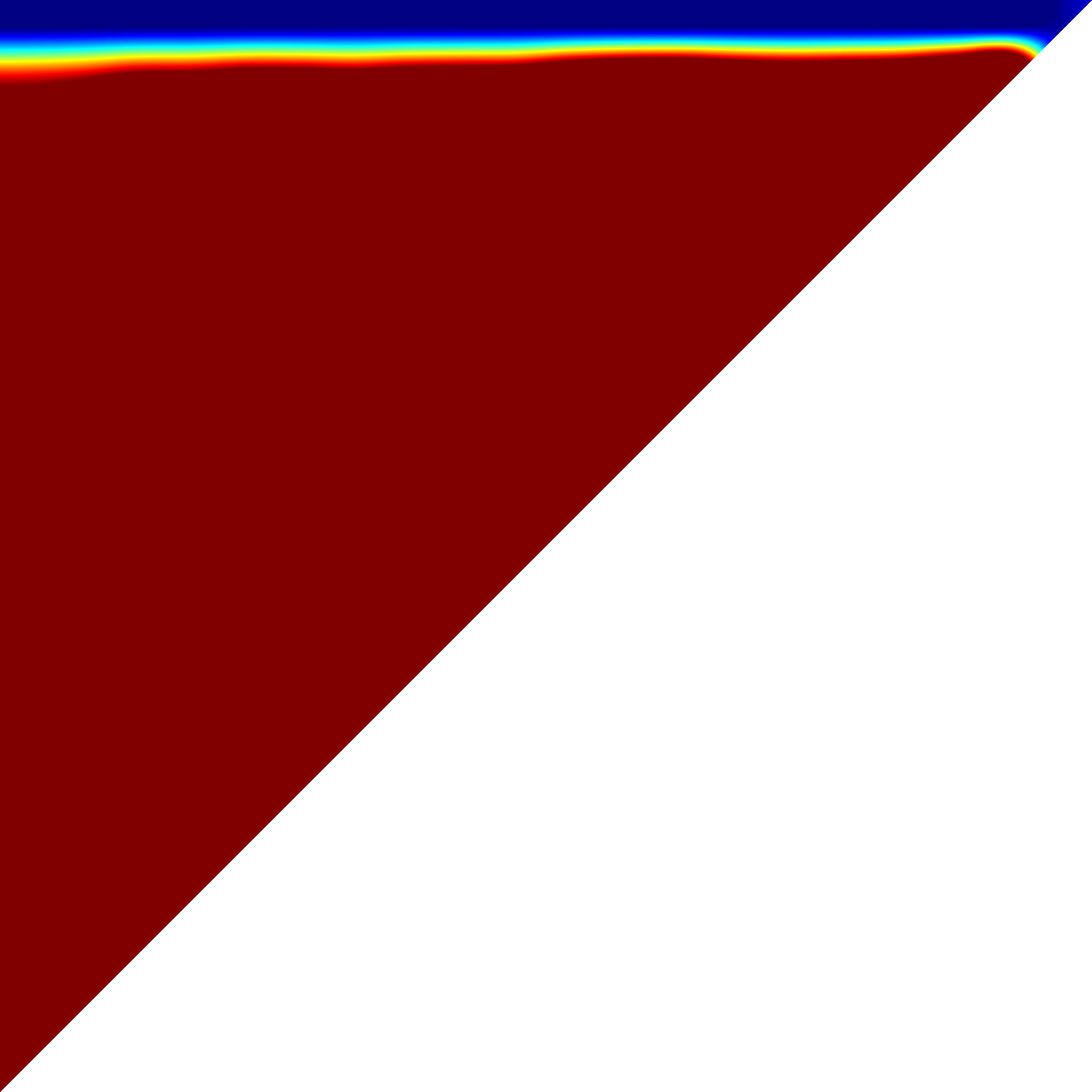};
\end{axis}
\end{tikzpicture}%

%% file: figures/A13_linear_CQT_clean.tex
%
%
\begin{tikzpicture}

\begin{axis}[%
width=.8in,
height=.8in,
at={(1.346in,0.502in)},
scale only axis,
point meta min=0,
point meta max=25,
axis on top,
xtick={0.5,500,1000,1500,2000},
xticklabels={{0},{2},{4},{6},{8}},
y dir=reverse,
xmin=.5,
xmax=2000.5,
ymin=.5,
ymax=2000.5,
ytick={0.5,1000},
yticklabels={},
axis background/.style={fill=white},
title style={align=center},
axis x line*=bottom,
axis y line*=left,
xtick style={draw=none},
ytick style={draw=none},
title={A13},
]
\addplot [forget plot] graphics [xmin=0.5, xmax=2000.5, ymin=0.5, ymax=2000.5] {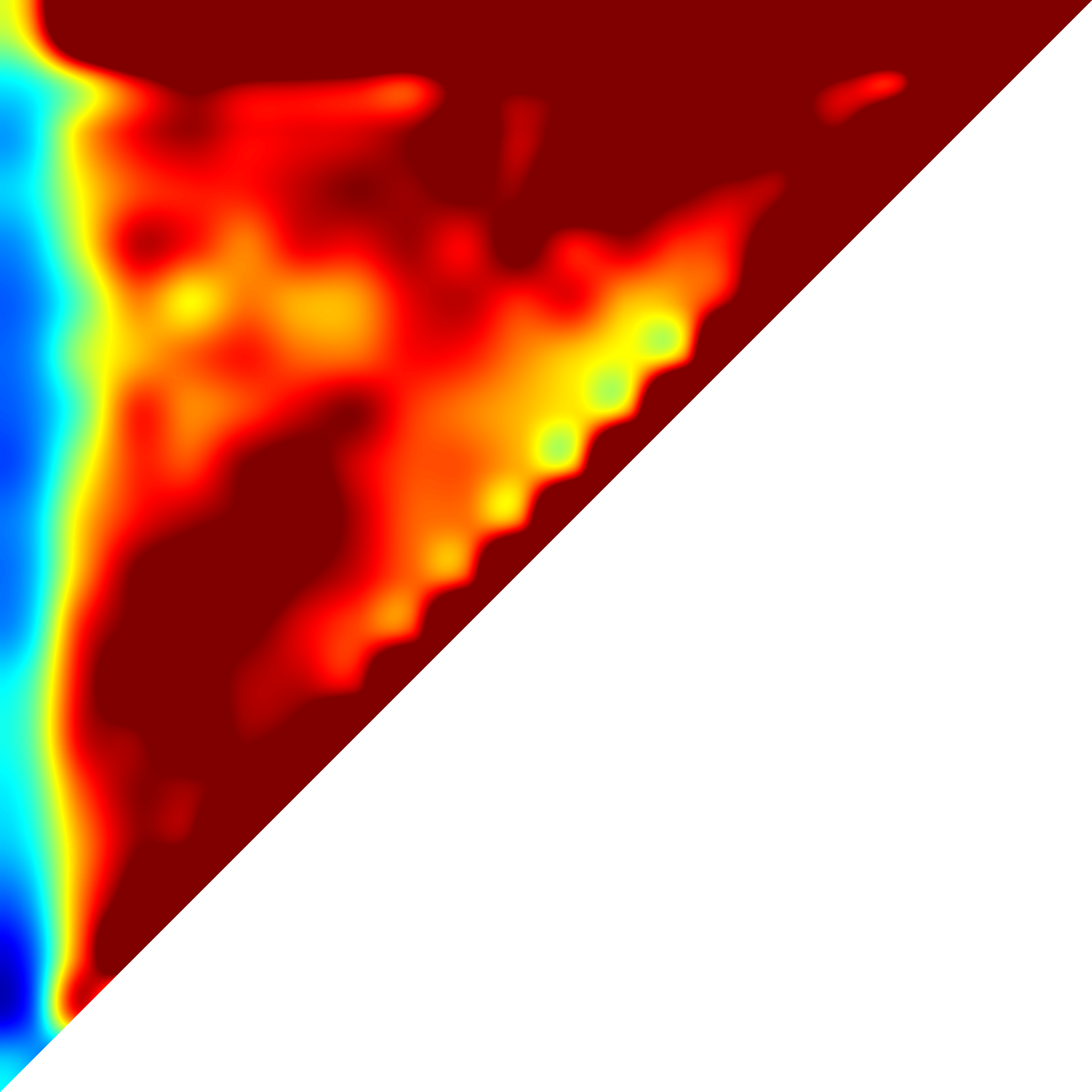};
\end{axis}
\end{tikzpicture}%

%% file: figures/A14_linear_CQT_clean.tex
%
%
\begin{tikzpicture}

\begin{axis}[%
width=.8in,
height=.8in,
at={(1.29in,0.502in)},
scale only axis,
point meta min=0,
point meta max=25,
axis on top,
xmin=0.5,
xmax=2000.5,
xtick={0.5,500,1000,1500,2000},
xticklabels={{0},{2},{4},{6},{8}},
y dir=reverse,
ymin=0.5,
ymax=2000.5,
ytick={},
yticklabels={},
axis background/.style={fill=white},
title={A14},
title style={align=center},
axis x line*=bottom,
axis y line*=left,
xtick style={draw=none},
ytick style={draw=none},
legend style={legend cell align=left, align=left, draw=white!15!black}
]
\addplot [forget plot] graphics [xmin=0.5, xmax=2000.5, ymin=0.5, ymax=2000.5] {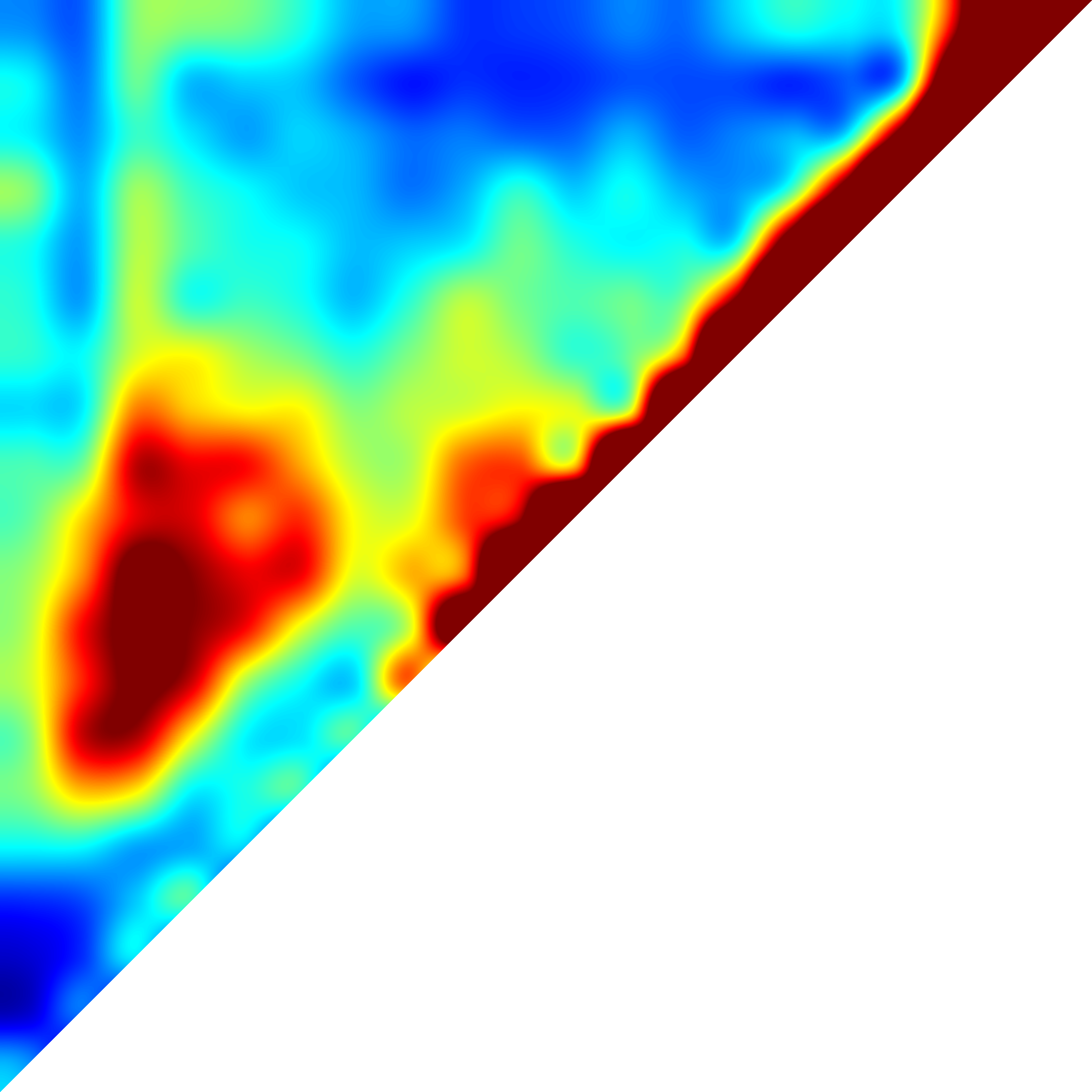};
\end{axis}
\end{tikzpicture}%

%% file: figures/A17_linear_CQT_clean.tex
%
%
\begin{tikzpicture}
\centering
\begin{axis}[%
width=.8in,
height=.8in,
at={(0.973in,0.502in)},
scale only axis,
point meta min=0,
point meta max=25,
axis on top,
xmin=0.5,
xmax=2000.5,
xtick={0.5,500,1000,1500,2000},
xticklabels={{0},{2},{4},{6},{8}},
y dir=reverse,
ymin=0.5,
ymax=2000.5,
ytick={0.5,500,1000,1500},
yticklabels={},
axis background/.style={fill=white},
title={A17},
title style={align=center},
axis x line*=bottom,
axis y line*=left,
xtick style={draw=none},
ytick style={draw=none},
legend style={legend cell align=left, align=left, draw=white!15!black},
colormap/jet,
colorbar,
colorbar style={ylabel style={font=\color{white!15!black}}, ylabel={EER (\%)}}
]
\addplot [forget plot] graphics [xmin=0.5, xmax=2000.5, ymin=0.5, ymax=2000.5] {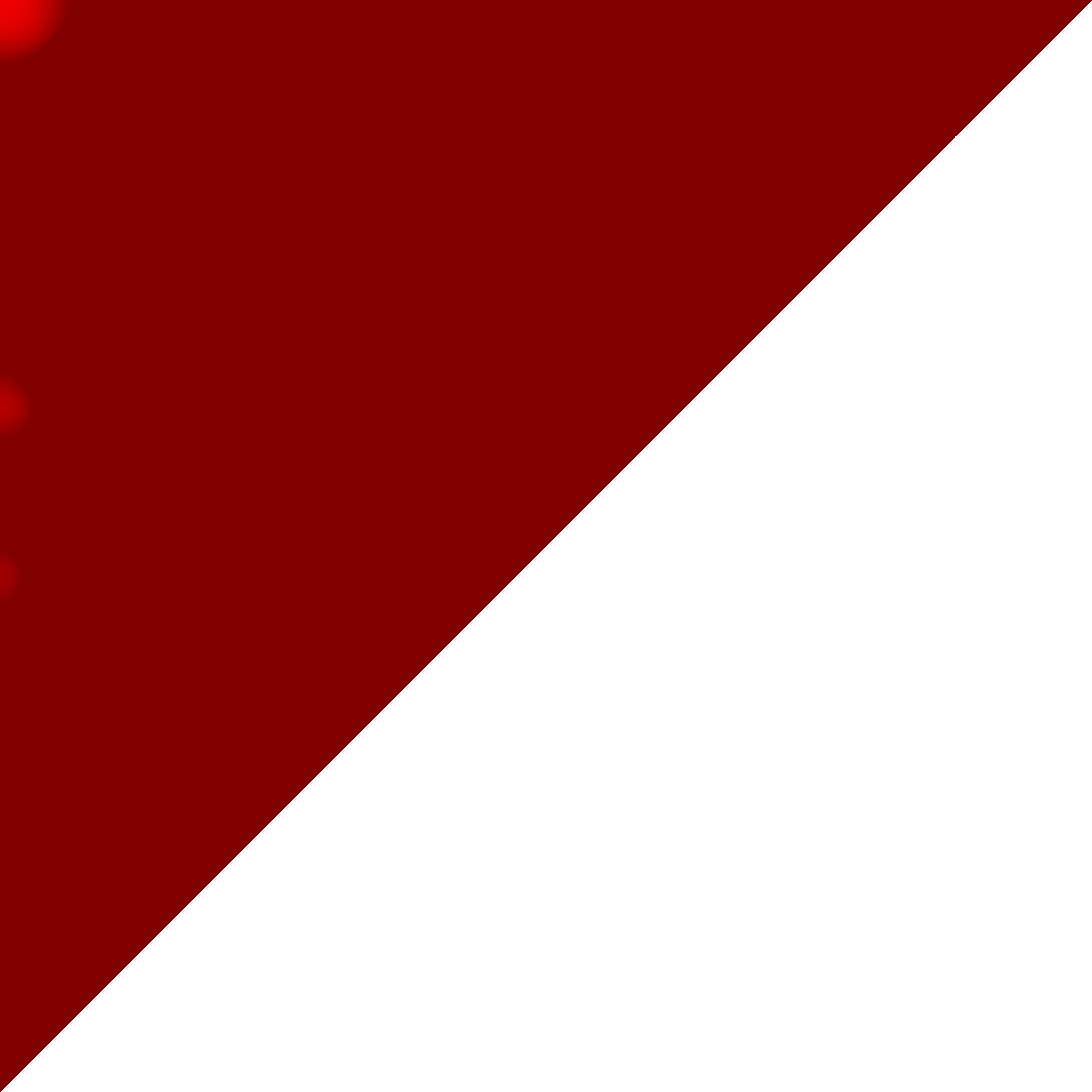};
\end{axis}
\end{tikzpicture}%

%% file: figures/A07_linear_FFT_clean.tex
%
%
\begin{tikzpicture}
\begin{axis}[%
width=.8in,
height=.8in,
scale only axis,
point meta min=0,
point meta max=25,
axis on top,
xmin=0.5,
xmax=2000.5,
xtick={0.5,500,1000,1500,2000},
xticklabels={{0},{2},{4},{6},{8}},
y dir=reverse,
ymin=0.5,
ymax=2000.5,
ytick={0.5,500,1000,1500},
yticklabels={{8},{6},{4},{2}},
ylabel style={font=\bfseries\color{white!15!black}},
ylabel={$\text{f}_{\text{max}}$} {\text{(kHz)}},
axis background/.style={fill=white},
title style={align=center},
axis x line*=bottom,
axis y line*=left,
xtick style={draw=none},
ytick style={draw=none}
]
\addplot [forget plot] graphics [xmin=0.5, xmax=2000.5, ymin=0.5, ymax=2000.5] {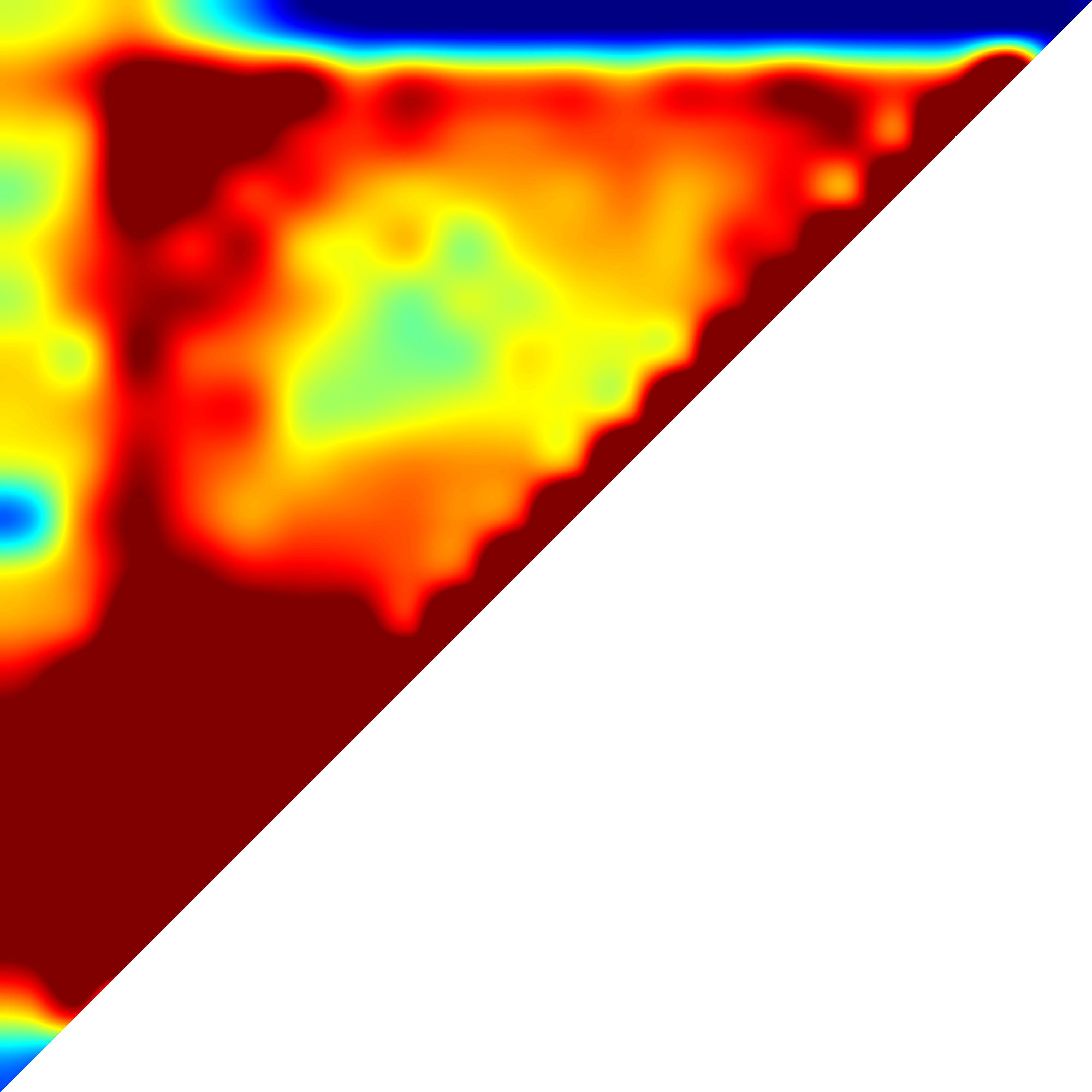};
\end{axis}
\end{tikzpicture}%

%% file: figures/A16_linear_FFT_clean.tex
%
%
\begin{tikzpicture}

\begin{axis}[%
width=.8in,
height=.8in,
at={(1.29in,0.502in)},
scale only axis,
point meta min=0,
point meta max=25,
axis on top,
xmin=0.5,
xmax=2000.5,
xtick={0.5,500,1000,1500,2000},
xticklabels={{0},{2},{4},{6},{8}},
y dir=reverse,
ymin=0.5,
ymax=2000.5,
ytick={0.5,500,1000,1500},
yticklabels={},
axis background/.style={fill=white},
axis x line*=bottom,
axis y line*=left,
xtick style={draw=none},
ytick style={draw=none}
]
\addplot [forget plot] graphics [xmin=0.5, xmax=2000.5, ymin=0.5, ymax=2000.5] {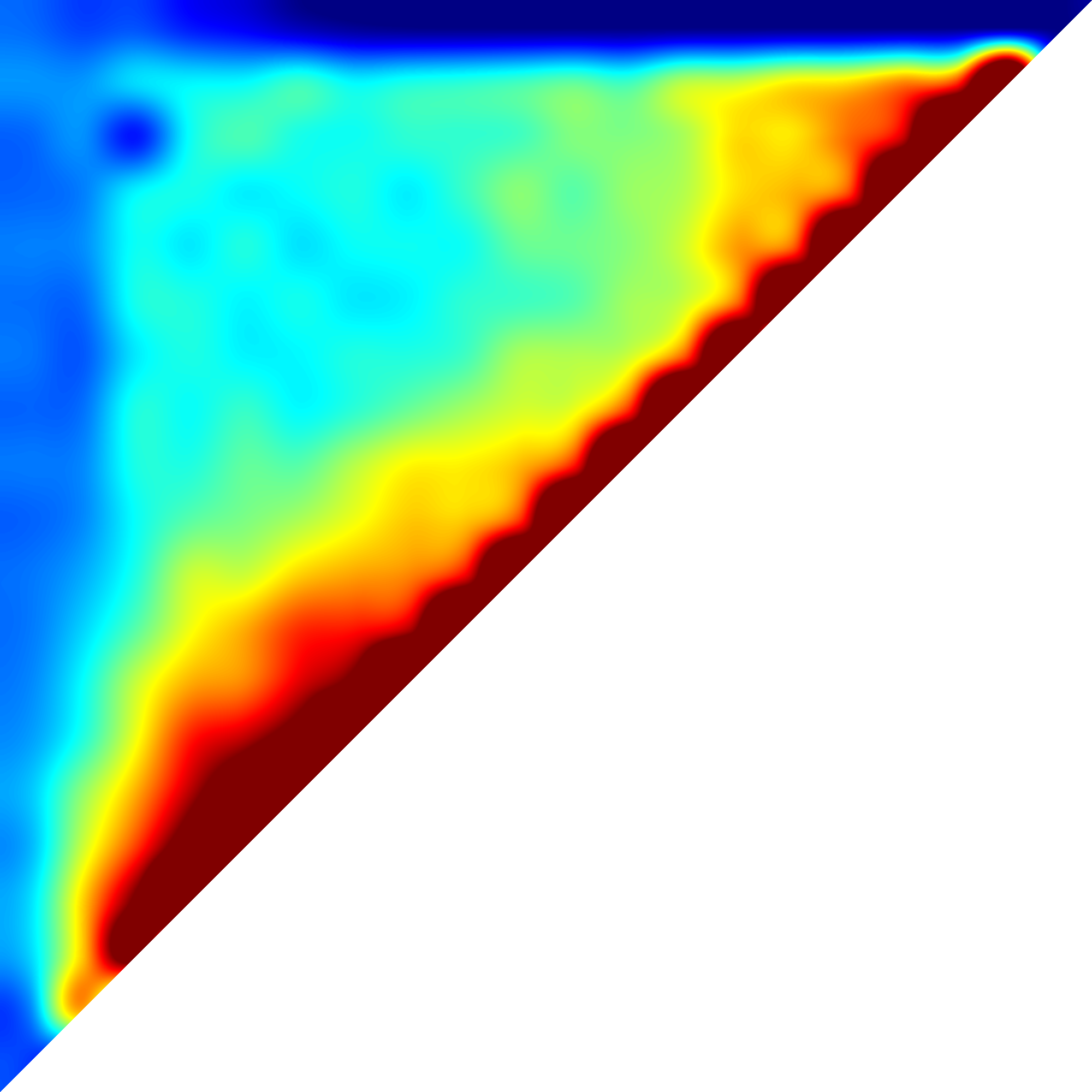};
\end{axis}
\end{tikzpicture}%

%% file: figures/A19_linear_FFT_clean.tex
%
%
\begin{tikzpicture}

\begin{axis}[%
width=.8in,
height=.8in,
at={(1.346in,0.502in)},
scale only axis,
point meta min=0,
point meta max=25,
axis on top,
xmin=0.5,
xmax=2000.5,
xtick={0.5,500,1000,1500,2000},
xticklabels={{0},{2},{4},{6},{8}},
y dir=reverse,
ymin=0.5,
ymax=2000.5,
ytick={},
yticklabels={},
axis background/.style={fill=white},
axis x line*=bottom,
axis y line*=left,
xtick style={draw=none},
ytick style={draw=none}
]
\addplot [forget plot] graphics [xmin=0.5, xmax=2000.5, ymin=0.5, ymax=2000.5] {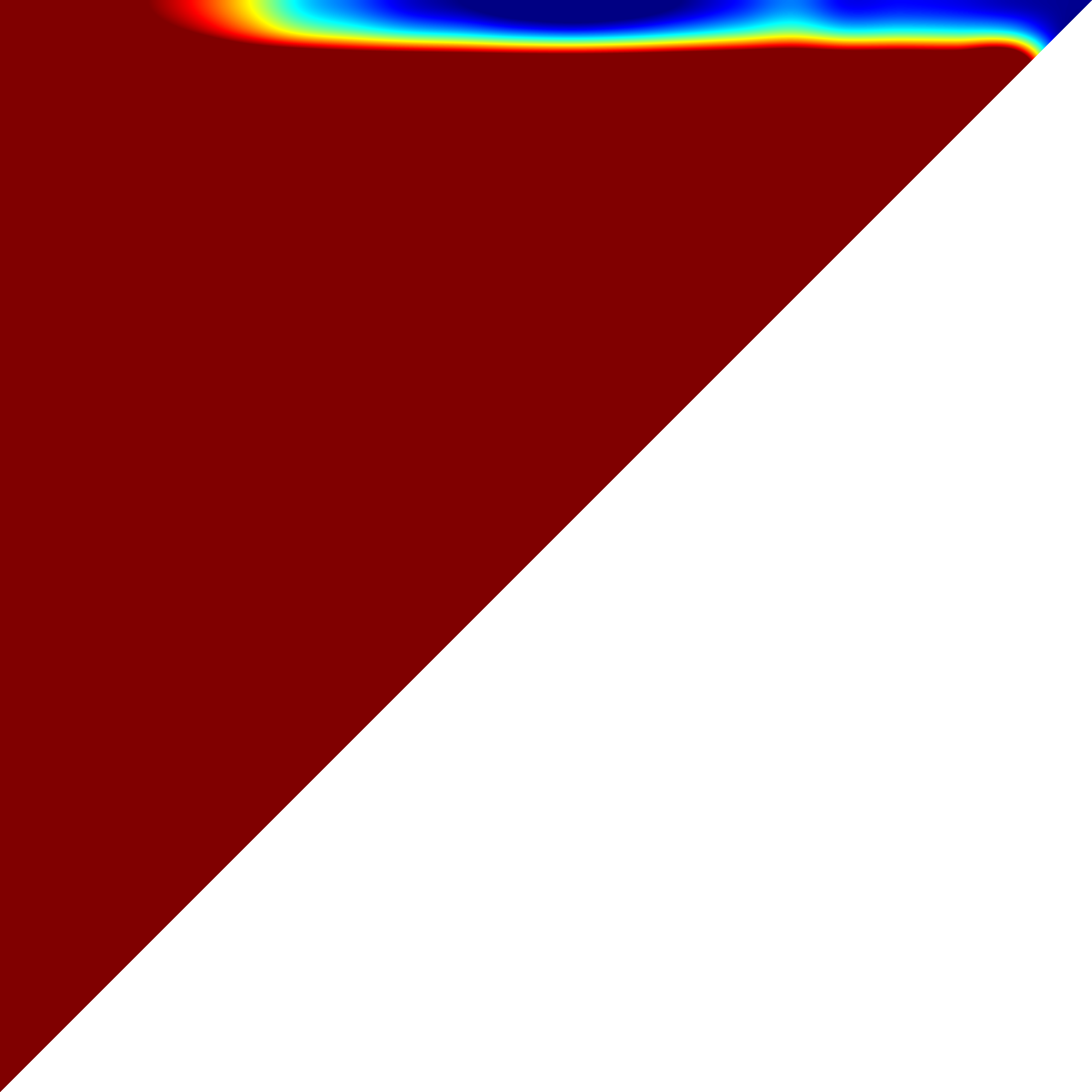};
\end{axis}

\end{tikzpicture}%

%% file: figures/A13_linear_FFT_clean.tex
%
%
\begin{tikzpicture}

\begin{axis}[%
width=.8in,
height=.8in,
scale only axis,
point meta min=0,
point meta max=25,
axis on top,
xmin=0.5,
xmax=2000.5,
xtick={0.5,500,1000,1500,2000},
xticklabels={{0},{2},{4},{6},{8}},
y dir=reverse,
ymin=0.5,
ymax=2000.5,
ytick={0.5,500,1000,1500},
yticklabels={},
axis background/.style={fill=white},
axis x line*=bottom,
axis y line*=left,
xtick style={draw=none},
ytick style={draw=none},
legend style={legend cell align=left, align=left, draw=white!15!black}
]
\addplot [forget plot] graphics [xmin=0.5, xmax=2000.5, ymin=0.5, ymax=2000.5] {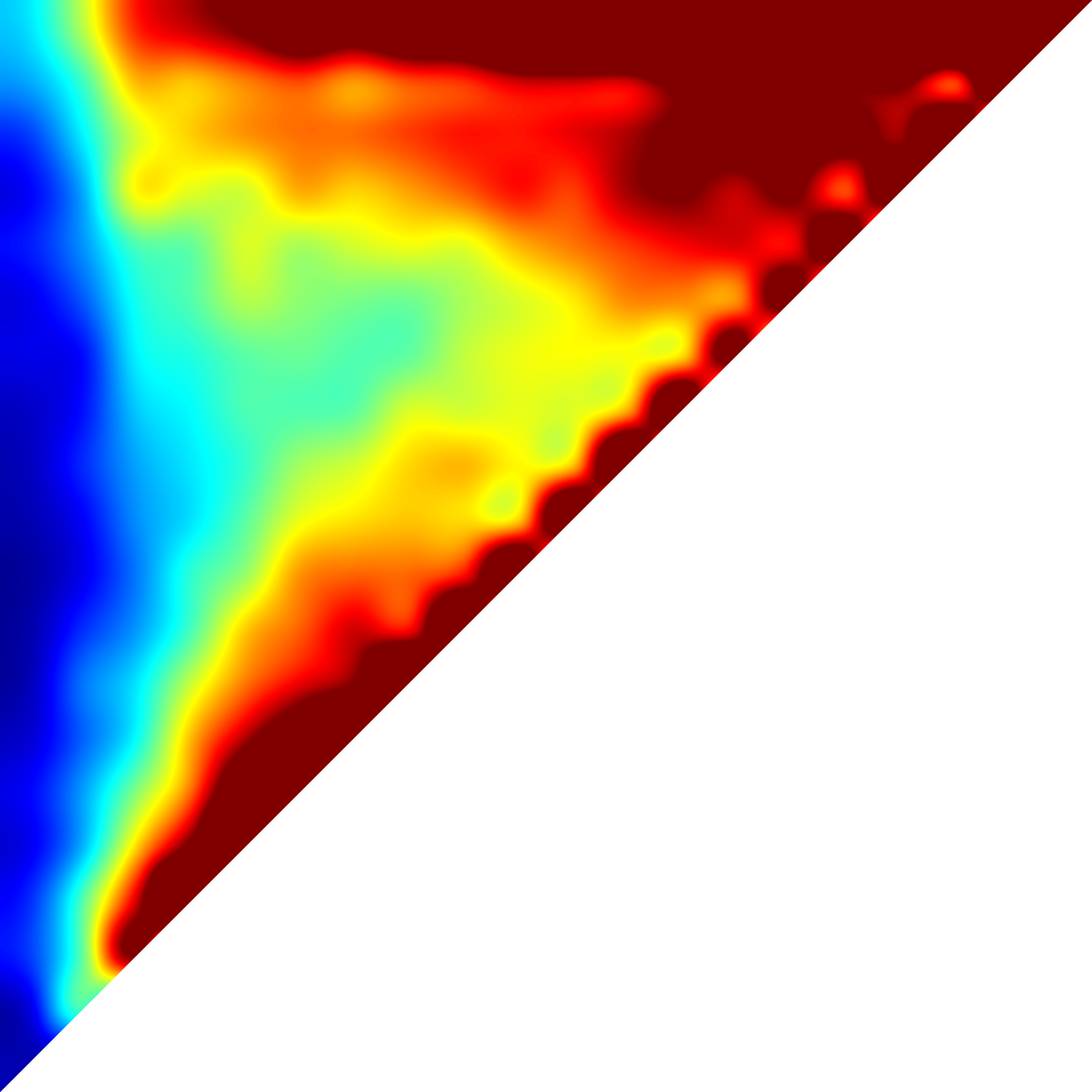};
\end{axis}

\end{tikzpicture}%

%% file: figures/A14_linear_FFT_clean.tex
%
%
\begin{tikzpicture}

\begin{axis}[%
width=.8in,
height=.8in,
at={(1.346in,0.502in)},
scale only axis,
point meta min=0,
point meta max=25,
axis on top,
xmin=0.5,
xmax=2000.5,
xtick={0.5,500,1000,1500,2000},
xticklabels={{0},{2},{4},{6},{8}},
y dir=reverse,
ymin=0.5,
ymax=2000.5,
ytick={0.5,500,1000,1500},
yticklabels={},
axis background/.style={fill=white},
axis x line*=bottom,
axis y line*=left,
xtick style={draw=none},
ytick style={draw=none}
]
\addplot [forget plot] graphics [xmin=0.5, xmax=2000.5, ymin=0.5, ymax=2000.5] {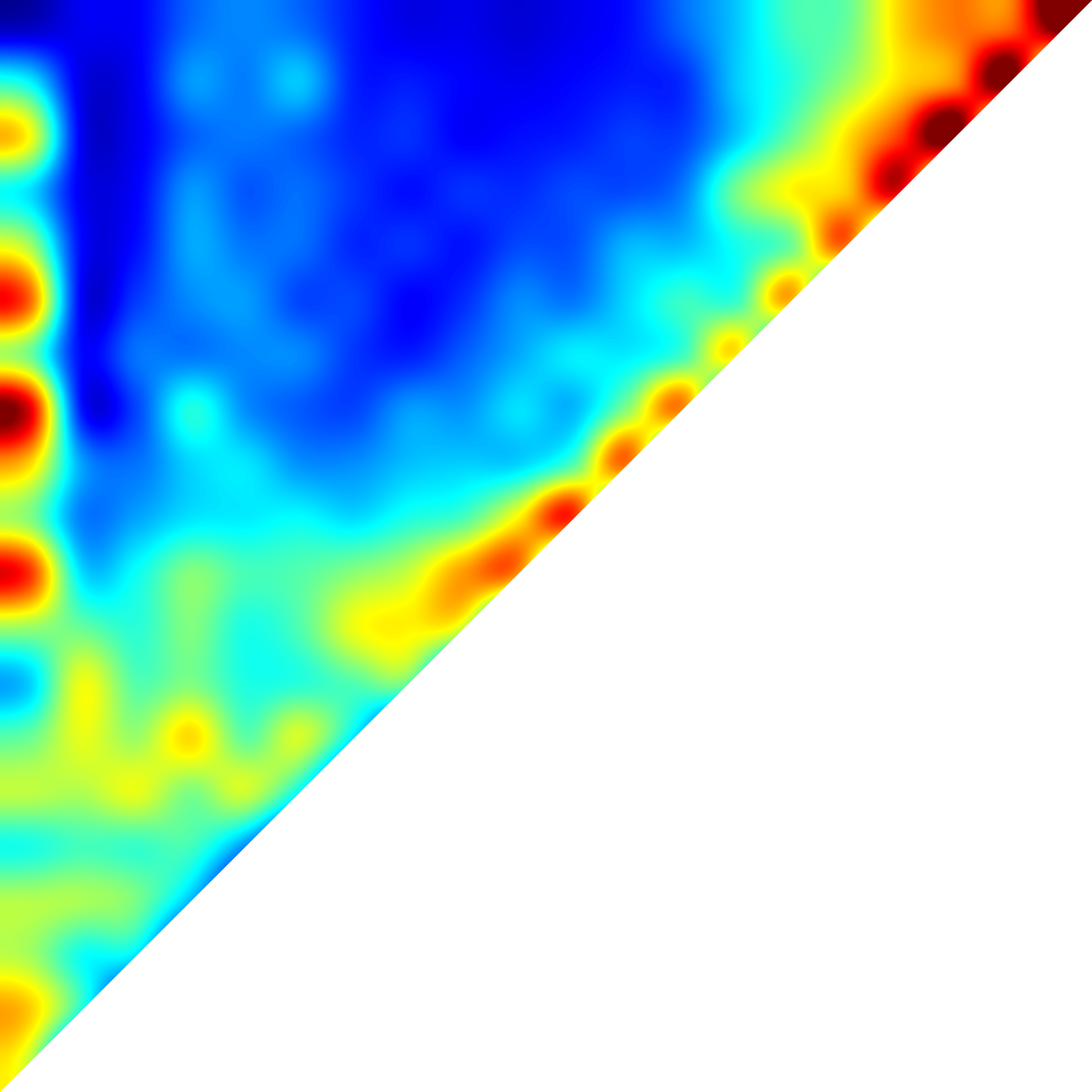};
\end{axis}
\end{tikzpicture}%

%% file: figures/A17_linear_FFT_clean.tex
%
%
\begin{tikzpicture}

\begin{axis}[%
width=.8in,
height=.8in,
at={(1.03in,0.502in)},
scale only axis,
point meta min=0,
point meta max=25,
axis on top,
xmin=0.5,
xmax=2000.5,
xtick={0.5,500,1000,1500,2000},
xticklabels={{0},{2},{4},{6},{8}},
y dir=reverse,
ymin=0.5,
ymax=2000.5,
ytick={0.5,500,1000,1500},
yticklabels={},
axis background/.style={fill=white},
axis x line*=bottom,
axis y line*=left,
xtick style={draw=none},
ytick style={draw=none},
legend style={legend cell align=left, align=left, draw=white!15!black},
colormap/jet,
colorbar,
colorbar style={ylabel style={font=\color{white!15!black}}, ylabel={EER (\%)}}
]
\addplot [forget plot] graphics [xmin=0.5, xmax=2000.5, ymin=0.5, ymax=2000.5] {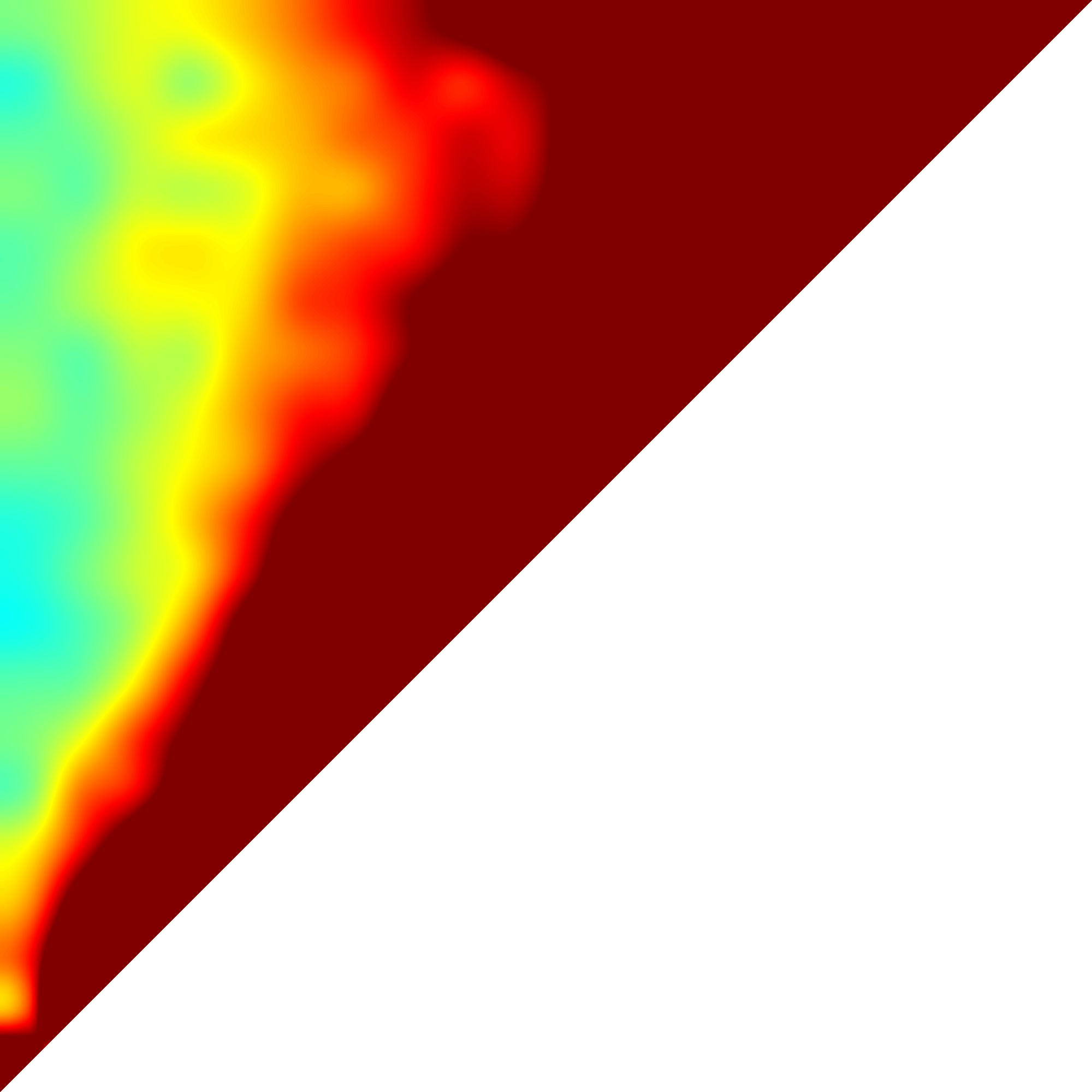};
\end{axis}
\end{tikzpicture}%

%% file: figures/A07_geometric_CQT_clean.tex
%
%
\begin{tikzpicture}

\begin{axis}[%
width=.8in,
height=.8in,
scale only axis,
point meta min=0,
point meta max=25,
axis on top,
xmin=0.5,
xmax=2000.5,
xtick={0.5,500,1000,1500,2000},
xticklabels={{0},{2},{4},{6},{8}},
xlabel style={font=\bfseries\color{white!15!black}},
xlabel={$\text{f}_{\text{min}}$} {\text{(kHz)}},
y dir=reverse,
ymin=0.5,
ymax=2000.5,
ytick={0.5,500,1000,1500},
yticklabels={{8},{6},{4},{2}},
ylabel style={font=\bfseries\color{white!15!black}},
ylabel={$\text{f}_{\text{max}}$} {\text{(kHz)}},
axis background/.style={fill=white},
axis x line*=bottom,
axis y line*=left,
xtick style={draw=none},
ytick style={draw=none}
]
\addplot [forget plot] graphics [xmin=0.5, xmax=2000.5, ymin=0.5, ymax=2000.5] {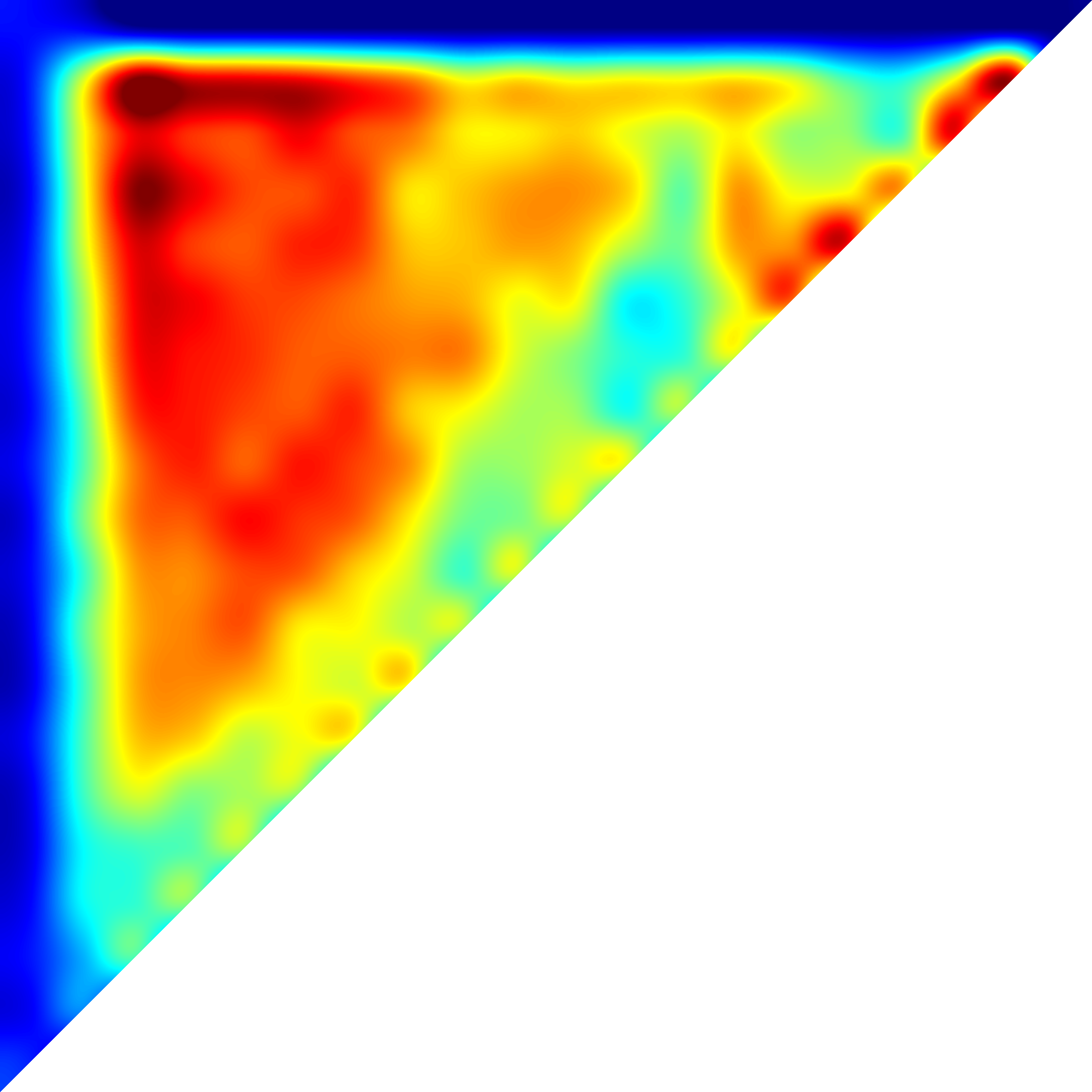};
\end{axis}

\end{tikzpicture}%

%% file: figures/A16_geometric_CQT_clean.tex
%
%
\begin{tikzpicture}

\begin{axis}[%
width=.8in,
height=.8in,
at={(1.373in,0.555in)},
scale only axis,
point meta min=0,
point meta max=25,
axis on top,
xmin=0.5,
xmax=2000.5,
xtick={0.5,500,1000,1500,2000},
xticklabels={{0},{2},{4},{6},{8}},
xlabel style={font=\bfseries\color{white!15!black}},
xlabel={$\text{f}_{\text{min}}$} {\text{(kHz)}},
y dir=reverse,
ymin=0.5,
ymax=2000.5,
ytick={0.5,500,1000,1500},
yticklabels={},
axis background/.style={fill=white},
axis x line*=bottom,
axis y line*=left,
xtick style={draw=none},
ytick style={draw=none}
]
\addplot [forget plot] graphics [xmin=0.5, xmax=2000.5, ymin=0.5, ymax=2000.5] {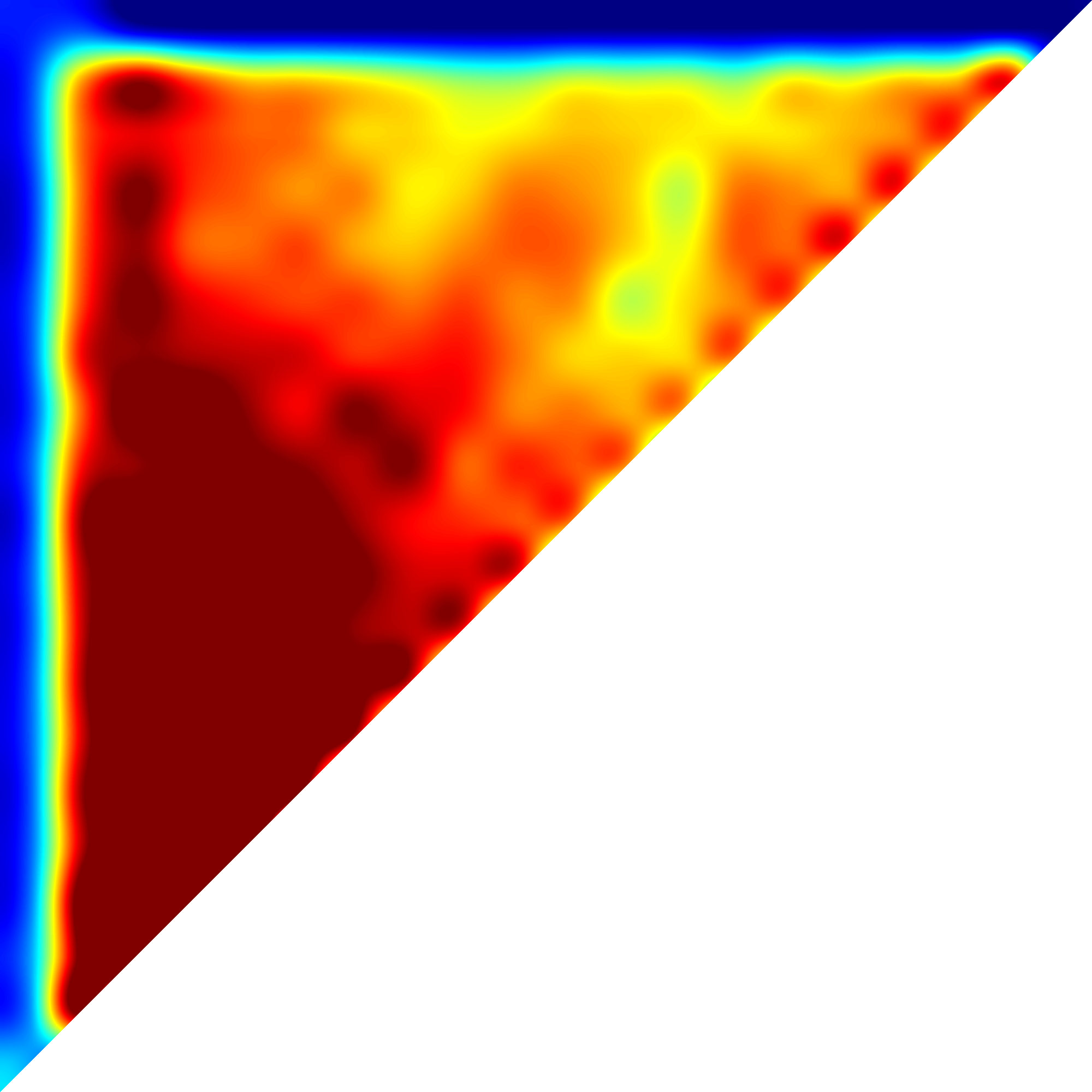};
\end{axis}
\end{tikzpicture}%

%% file: figures/A19_geometric_CQT_clean.tex
%
%
\begin{tikzpicture}

\begin{axis}[%
width=.8in,
height=.8in,
scale only axis,
point meta min=0,
point meta max=25,
axis on top,
xmin=0.5,
xmax=2000.5,
xtick={0.5,500,1000,1500,2000},
xticklabels={{0},{2},{4},{6},{8}},
xlabel style={font=\bfseries\color{white!15!black}},
xlabel={$\text{f}_{\text{min}}$} {\text{(kHz)}},
y dir=reverse,
ymin=0.5,
ymax=2000.5,
ytick={},
yticklabels={},
axis background/.style={fill=white},
axis x line*=bottom,
axis y line*=left,
xtick style={draw=none},
ytick style={draw=none}
]
\addplot [forget plot] graphics [xmin=0.5, xmax=2000.5, ymin=0.5, ymax=2000.5] {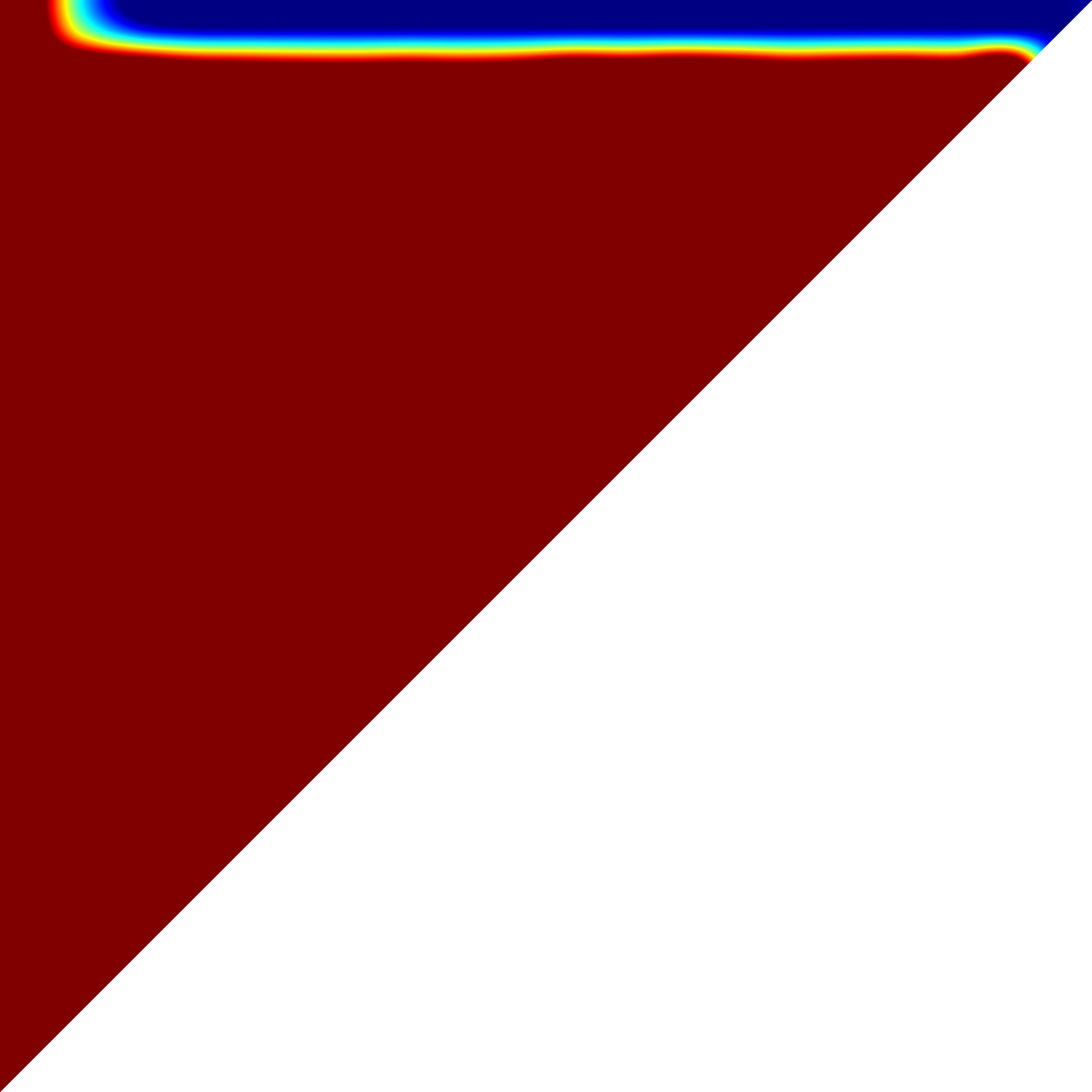};
\end{axis}

\end{tikzpicture}%

%% file: figures/A13_geometric_CQT_clean.tex
%
%
\begin{tikzpicture}

\begin{axis}[%
width=.8in,
height=.8in,
scale only axis,
point meta min=0,
point meta max=25,
axis on top,
xmin=0.5,
xmax=2000.5,
xtick={0.5,500,1000,1500,2000},
xticklabels={{0},{2},{4},{6},{8}},
xlabel style={font=\bfseries\color{white!15!black}},
xlabel={$\text{f}_{\text{min}}$} {\text{(kHz)}},
y dir=reverse,
ymin=0.5,
ymax=2000.5,
ytick={0.5,500,1000,1500},
yticklabels={},
axis background/.style={fill=white},
axis x line*=bottom,
axis y line*=left,
xtick style={draw=none},
ytick style={draw=none},
]
\addplot [forget plot] graphics [xmin=0.5, xmax=2000.5, ymin=0.5, ymax=2000.5] {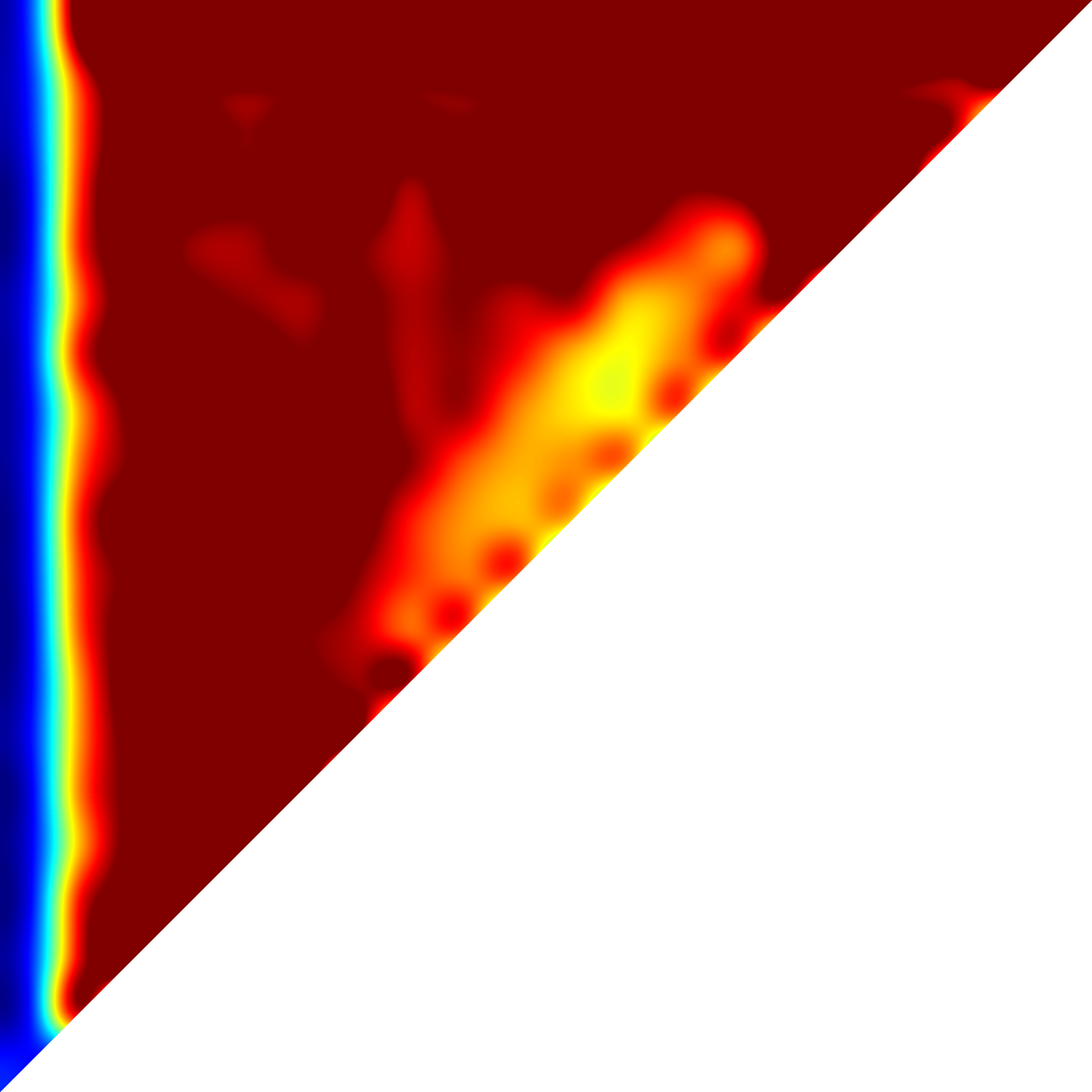};
\end{axis}
\end{tikzpicture}%

%% file: figures/A14_geometric_CQT_clean.tex
%
%
\begin{tikzpicture}

\begin{axis}[%
width=.8in,
height=.8in,
scale only axis,
point meta min=0,
point meta max=25,
axis on top,
xmin=0.5,
xmax=2000.5,
xtick={0.5,500,1000,1500,2000},
xticklabels={{0},{2},{4},{6},{8}},
xlabel style={font=\bfseries\color{white!15!black}},
xlabel={$\text{f}_{\text{min}}$} {\text{(kHz)}},
y dir=reverse,
ymin=0.5,
ymax=2000.5,
ytick={0.5,500,1000,1500},
yticklabels={},
axis background/.style={fill=white},
axis x line*=bottom,
axis y line*=left,
xtick style={draw=none},
ytick style={draw=none}
]
\addplot [forget plot] graphics [xmin=0.5, xmax=2000.5, ymin=0.5, ymax=2000.5] {figures/A14_geometric_CQT_clean-1.png};
\end{axis}
\end{tikzpicture}%

%% file: figures/A17_geometric_CQT_clean.tex
%
%
\begin{tikzpicture}

\begin{axis}[%
width=.8in,
height=.8in,
scale only axis,
point meta min=0,
point meta max=25,
axis on top,
xmin=0.5,
xmax=2000.5,
xtick={0.5,500,1000,1500,2000},
xticklabels={{0},{2},{4},{6},{8}},
xlabel style={font=\bfseries\color{white!15!black}},
xlabel={$\text{f}_{\text{min}}$} {\text{(kHz)}},
y dir=reverse,
ymin=0.5,
ymax=2000.5,
ytick={0.5,500,1000,1500},
yticklabels={},
axis background/.style={fill=white},
axis x line*=bottom,
axis y line*=left,
xtick style={draw=none},
ytick style={draw=none},
colormap/jet,
colorbar,
colorbar style={ylabel style={font=\color{white!15!black}}, ylabel={EER (\%)}}
]
\addplot [forget plot] graphics [xmin=0.5, xmax=2000.5, ymin=0.5, ymax=2000.5] {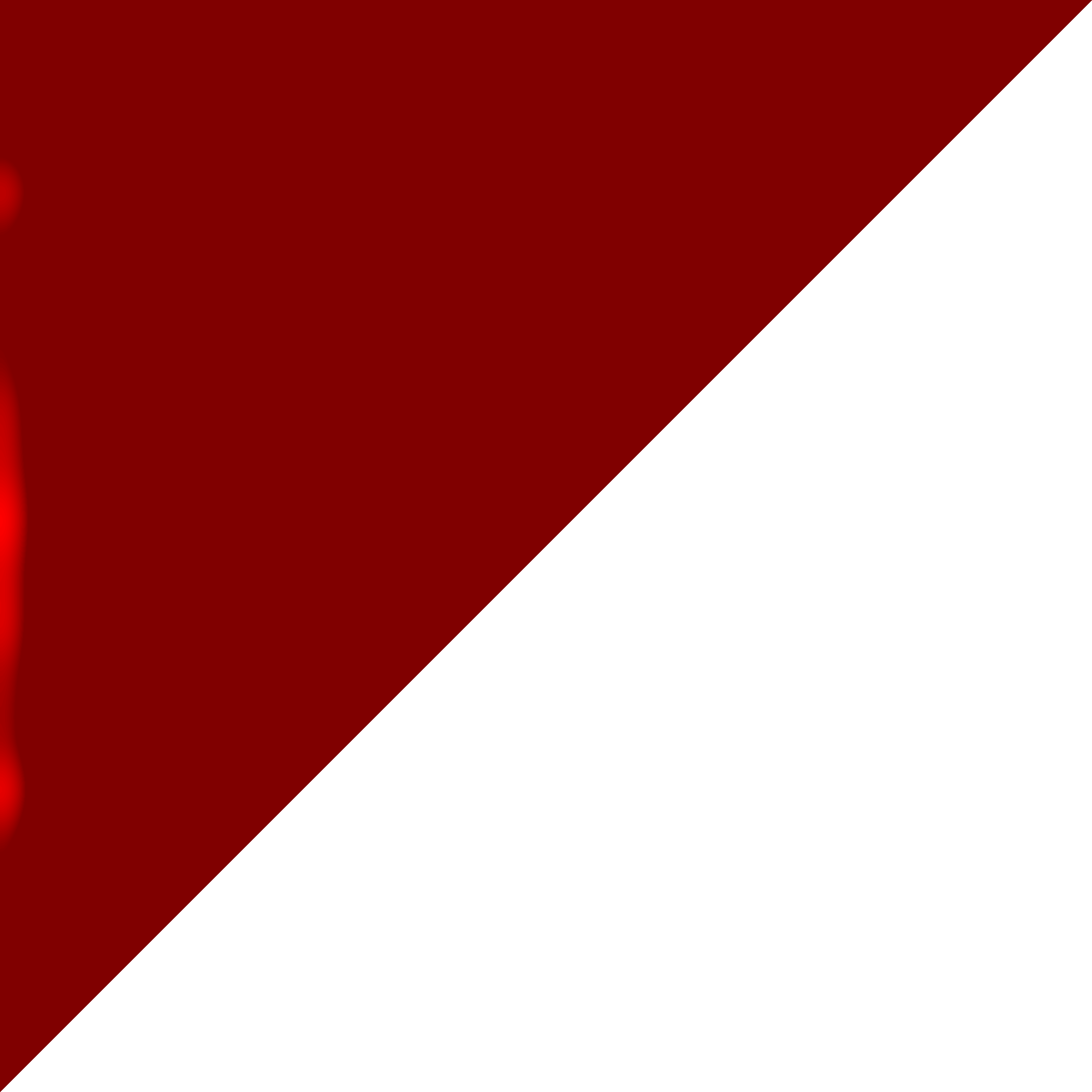};
\end{axis}
\end{tikzpicture}%

%% file: main_arxiv.bbl
\begin{thebibliography}{10}

\bibitem{hansen2015speaker}
J.~Hansen and T.~Hasan,
\newblock ``Speaker recognition by machines and humans: A tutorial review,''
\newblock {\em in IEEE Signal Processing Magazine}, vol. 32, no. 6, pp. 74--99,
  2015.

\bibitem{kinnunen2010overview}
T.~Kinnunen and H.~Li,
\newblock ``An overview of text-independent speaker recognition: From features
  to supervectors,''
\newblock {\em Speech communication}, vol. 52, no. 1, pp. 12--40, 2010.

\bibitem{EURECOM+4018}
N.~{E}vans, T.~{K}innunen, and J.~{Y}amagishi,
\newblock ``{S}poofing and countermeasures for automatic speaker
  verification,''
\newblock in {\em {Proc. INTERSPEECH}}, {L}yon, {FRANCE}, 2013, pp. 925--929.

\bibitem{evans2013spoofing}
N.~Evans, J.~Yamagishi, and T.~Kinnunen,
\newblock ``Spoofing and countermeasures for speaker verification: a need for
  standard corpora, protocols and metrics,''
\newblock {\em IEEE Signal Processing Society Speech and Language Technical
  Committee Newsletter}, 2013.

\bibitem{wu2015asvspoof}
Z.~Wu, T.~Kinnunen, N.~Evans, J.~Yamagishi, C.~Hanilci, M.~Sahidullah, and
  A.~Sizov,
\newblock ``{ASV}spoof 2015: the first automatic speaker verification spoofing
  and countermeasures challenge,''
\newblock in {\em Proc. INTERSPEECH}, Dresden, Germany, 2015, pp. 2037--2041.

\bibitem{todisco2016new}
M.~Todisco, H.~Delgado, and N.~Evans,
\newblock ``A new feature for automatic speaker verification anti-spoofing:
  Constant {Q} cepstral coefficients,''
\newblock in {\em Proc. Speaker Odyssey Workshop}, Bilbao, Spain, 2016, pp.
  249--252.

\bibitem{todisco2017constant}
M.~Todisco, H.~Delgado, and N.~Evans,
\newblock ``{Constant Q cepstral coefficients: A spoofing countermeasure for
  automatic speaker verification},''
\newblock {\em Computer Speech \& Language}, vol. 45, pp. 516--535, 2017.

\bibitem{chettri2019ensemble}
B.~Chettri, D.~Stoller, V.~Morfi, M.~A~Mart{\'\i}nez Ram{\'\i}rez, E.~Benetos,
  and B.~L Sturm,
\newblock ``Ensemble models for spoofing detection in automatic speaker
  verification,''
\newblock in {\em Proc. INTERSPEECH}, Graz, Austria, 2019, pp. 1118--1112.

\bibitem{alluri2019iiit}
R.~Alluri and A.~Vuppala,
\newblock ``{IIIT}-{H} spoofing countermeasures for automatic speaker
  verification spoofing and countermeasures challenge 2019,''
\newblock in {\em Proc. INTERSPEECH}, Graz, Austria, 2019, pp. 1043--1047.

\bibitem{yang2019sjtu}
Y.~Yang, H.~Wang, H.~Dinkel, Z.~Chen, S.~Wang, Y.~Qian, and K.~Yu,
\newblock ``The {SJTU} robust anti-spoofing system for the {ASV}spoof 2019
  challenge,''
\newblock in {\em Proc. INTERSPEECH}, Graz, Austria, 2019, pp. 1038--1042.

\bibitem{lavrentyeva2019stc}
G.~Lavrentyeva, S.~Novoselov, A.~Tseren, M.~Volkova, A.~Gorlanov, and
  A.~Kozlov,
\newblock ``{STC} antispoofing systems for the {ASV}spoof2019 challenge,''
\newblock in {\em Proc. INTERSPEECH}, Graz, Austria, 2019, pp. 1033--1037.

\bibitem{saito2011one}
D.~Saito, K.~Yamamoto, N.~Minematsu, and K.~Hirose,
\newblock ``One-to-many voice conversion based on tensor representation of
  speaker space,''
\newblock in {\em Proc. INTERSPEECH}, Florence, Italy, 2011, pp. 653--656.

\bibitem{kinnunen2016utterance}
T.~Kinnunen, M.~Sahidullah, I.~Kukanov, H.~Delgado, M.~Todisco, A.~Sarkar, N.~B
  Thomsen, V.~Hautam{\"a}ki, N.~Evans, and Z.~Hua Tan,
\newblock ``Utterance verification for text-dependent speaker recognition: a
  comparative assessment using the {RedDots} corpus,''
\newblock in {\em Proc. INTERSPEECH}, San Francisco, USA, 2016, pp. 430--434.

\bibitem{schorkhuber2014matlab}
C.~Sch{\"o}rkhuber, A.~Klapuri, N.~Holighaus, and M.~D{\"o}rfler,
\newblock ``A matlab toolbox for efficient perfect reconstruction
  time-frequency transforms with log-frequency resolution,''
\newblock in {\em Proc. Audio Engineering Society International Conference on
  Semantic Audio}, London, UK, 2014.

\bibitem{Youngberg78}
J.~Youngberg and S.~Boll,
\newblock ``Constant-q signal analysis and synthesis,''
\newblock in {\em IEEE International Conference on Acoustics, Speech, and
  Signal Processing (ICASSP)}, Tulsa, Oklahoma, USA, 1978, vol.~3, pp.
  375--378.

\bibitem{Brown91}
J.~Brown,
\newblock ``Calculation of a constant {Q} spectral transform,''
\newblock {\em Journal of the Acoustical Society of America (JASA)}, vol. 89,
  no. 1, pp. 425--434, 1991.

\bibitem{Velasco11}
G.~A Velasco, N.~Holighaus, M.~Dorfler, and T.~Grill,
\newblock ``Constructing an invertible constant-{Q} transform with
  nonstationary {G}abor frames,''
\newblock in {\em Proc. Digital Audio Effects (DAFx-11)}, Paris, France, 2011,
  pp. 93--99.

\bibitem{quatieri2006discrete}
T.~F Quatieri,
\newblock {\em Discrete-time speech signal processing: principles and
  practice},
\newblock Pearson Education, 2006.

\bibitem{deller1993discrete}
J.~R Deller~Jr, J.~G Proakis, and J.~H Hansen,
\newblock {\em Discrete time processing of speech signals},
\newblock Prentice Hall PTR, 1993.

\bibitem{sriskandaraja2016investigation}
K.~Sriskandaraja, V.~Sethu, P.~Ngoc Le, and E.~Ambikairajah,
\newblock ``Investigation of sub-band discriminative information between
  spoofed and genuine speech,''
\newblock in {\em Proc. INTERSPEECH}, San Francisco, USA, 2016, pp. 1710--1714.

\bibitem{yang2019significance}
J.~Yang, R.~K Das, and H.~Li,
\newblock ``Significance of subband features for synthetic speech detection,''
\newblock {\em IEEE Transactions on Information Forensics and Security}, 2019.

\bibitem{brummer2013bosaris}
N.~Br{\"u}mmer and E.~De~Villiers,
\newblock ``The {bosaris} toolkit: Theory, algorithms and code for surviving
  the new dcf,''
\newblock {\em arXiv preprint arXiv:1304.2865}, 2013.

\bibitem{todisco2019asvspoof}
M.~Todisco, X.~Wang, V.~Vestman, M.~Sahidullah, H.~Delgado, A.~Nautsch,
  J.~Yamagishi, N.~Evans, T.~Kinnunen, and K.~Lee,
\newblock ``{ASV}spoof 2019: Future horizons in spoofed and fake audio
  detection,''
\newblock in {\em Proc. INTERSPEECH}, Graz, Austria, 2019, pp. 1008--1012.

\bibitem{wang2019asvspoof}
X.~Wang, J.~Yamagishi, M.~Todisco, H.~Delgado, A.~Nautsch, N.~Evans,
  M.~Sahidullah, V.~Vestman, T.~Kinnunen, K.~Lee, et~al.,
\newblock ``The {ASV}spoof 2019 database,''
\newblock {\em arXiv preprint arXiv:1911.01601}, 2019.

\bibitem{wavecyclegan}
K.~Tanaka, H.~Kameoka, and N.~Kaneko, T.and~Hojo,
\newblock ``Wavecyclegan2: Time-domain neural post-filter for speech waveform
  generation,''
\newblock {\em arXiv preprint arXiv:1904.02892}, 2019.

\bibitem{wu2016merlin}
Z.~Wu, O.~Watts, and S.~King,
\newblock ``Merlin: An open source neural network speech synthesis system.,''
\newblock in {\em Speech synthesis workshop ({SSW})}, Sunnyvale, USA, 2016, pp.
  202--207.

\bibitem{li2015generative}
Y.~Li, K.~Swersky, and R.~Zemel,
\newblock ``Generative moment matching networks,''
\newblock in {\em International Conference on Machine Learning}, Lille, France,
  2015, pp. 1718--1727.

\bibitem{kobayashi2018intra}
K.~Kobayashi, T.~Toda, and S.~Nakamura,
\newblock ``Intra-gender statistical singing voice conversion with direct
  waveform modification using log-spectral differential,''
\newblock {\em Speech Communication}, vol. 99, pp. 211--220, 2018.

\bibitem{ljliu2018wav}
L.~Juan Liu, Z.~Hua Ling, Y.~Jiang, M.~Zhou, and L.~Rong Dai,
\newblock ``{WaveNet} vocoder with limited training data for voice
  conversion,''
\newblock in {\em Proc. INTERSPEECH}, Hydrabad, India, 2018, pp. 1983--1987.

\bibitem{Kawahara1999Restructuring}
H.~Kawahara, I.~Masuda-Katsuse, and A.~De Cheveign{\'e},
\newblock ``Restructuring speech representations using a pitch-adaptive
  {time-frequency} smoothing and an {instantaneous-frequency} based {F0}
  extraction: Possible role of a repetitive structure in sounds,''
\newblock {\em Speech Communication}, vol. 27, no. 3--4, pp. 187--207, 1999.

\bibitem{McAuliffe/etal:2017:IS}
M.~McAuliffe, M.~Socolof, S.~Mihuc, M.~Wagner, and M.~Sonderegger,
\newblock ``{M}ontreal {F}orced {A}ligner: Trainable text-speech alignment
  using {K}aldi,''
\newblock in {\em Proc. INTERSPEECH}, Stockholm, Sweden, 2017, pp. 498--502.

\bibitem{2019arXiv190711898H}
W.~Chin {Huang}, Y.~Chiao {Wu}, K.~{Kobayashi}, Y.~Huai {Peng}, H.~Te {Hwang},
  P.~{Lumban Tobing}, Y.~{Tsao}, H.~Min {Wang}, and T.~{Toda},
\newblock ``Generalization of spectrum differential based direct waveform
  modification for voice conversion,''
\newblock in {\em Proc. Speech synthesis workshop (SSW)}, Vienna, Austria,
  2019, pp. 57--62.

\bibitem{kinnunen2018t}
T.~Kinnunen, K.~Lee, H.~Delgado, N.~Evans, M.~Todisco, J.~Sahidullah,
  M.and~Yamagishi, and D.~A Reynolds,
\newblock ``{t-DCF: a detection cost function for the tandem assessment of
  spoofing countermeasures and automatic speaker verification},''
\newblock in {\em {Proc. Speaker Odyssey Workshop}}, Les Sables d'Olonne,
  France, 2018, pp. 312--319.

\bibitem{brown1999computer}
J.~Brown,
\newblock ``Computer identification of musical instruments using pattern
  recognition with cepstral coefficients as features,''
\newblock {\em The Journal of the Acoustical Society of America (JASA)}, vol.
  105, no. 3, pp. 1933--1941, 1999.

\end{thebibliography}
